\documentclass[twoside,twocolumn,english,aps,showpacs,pra,superscriptaddress]{revtex4-1}

	\usepackage[usenames,dvipsnames]{xcolor}
	\usepackage{amsmath}
	\usepackage{amsfonts}
	\usepackage{amssymb}
	\usepackage[colorlinks=true,citecolor=blue,linkcolor=red]{hyperref}
	\usepackage{float}
	\usepackage{graphicx}
	\usepackage{subfigure}
	\usepackage{bbold}					
	\usepackage[makeroom]{cancel}		
	\usepackage{multirow}				
	\usepackage[normalem]{ulem}        
	\usepackage{array}
	\usepackage{pdfpages}
	\usepackage{braket}

	\makeatletter
	\AtBeginDocument{\let\LS@rot\@undefined}
	\makeatother
	
	\newcolumntype{x}[1]{>{\centering\let\newline\\\arraybackslash\hspace{0pt}}p{#1}}
	
	\renewcommand{\Im}{\operatorname{Im}}

	\DeclareMathAlphabet{\mathbbold}{U}{bbold}{m}{n}
	\def\imi{\mathrm{i}}				
	\def\e#1{\mathrm{e}^{#1}}				

	\DeclareMathOperator{\Tr}{Tr}
	\DeclareMathOperator{\Ker}{Ker}
	
	\newcounter{subeqn} %
	\makeatletter
	\@addtoreset{subeqn}{equation}
	\makeatother

\definecolor{TB}{rgb}{0,0,0} 

\begin{document}
\title{An analytic study of the independent coherent errors in the surface code}

\author{Yuanchen Zhao}
\affiliation{State Key Laboratory of Low Dimensional Quantum Physics, Department of Physics, Tsinghua University, Beijing, 100084, China}

\author{Dong E. Liu}
\email{dongeliu@mail.tsinghua.edu.cn}
\affiliation{State Key Laboratory of Low Dimensional Quantum Physics, Department of Physics, Tsinghua University, Beijing, 100084, China}
\affiliation{Beijing Academy of Quantum Information Sciences, Beijing 100193, China}
\affiliation{Frontier Science Center for Quantum Information, Beijing 100184, China}

\date{\today}

\begin{abstract}
The realistic coherent errors could induce very different behaviors compared with their stochastic counterparts in the quantum error correction (QEC) and fault tolerant quantum computation. Their impacts are believed to be very subtle, more detrimental and hard to analyze compared to those ideal stochastic errors. In this paper, we study the independent coherent error due to the imperfect unitary rotation on each physical qubit of the toric code. We find that the surface code under coherent error satisfies generalized Knill-Laflamme (K-L) criterion and falls into the category of approximate QEC. The extra term in the generalized K-L criterion corresponds to the coherent part of the error channel at logical level, and then show that the generalized K-L criterion approaches the normal K-L criterion when the code distance becomes large. In addition, we also find that if the code with a fixed distance $d$ is $\epsilon$- correctable, the value of $\epsilon$ describing the accuracy of the approximate QEC cannot be smaller than a lower bound. We then study the success probability of QEC under such coherent errors, and confirm that the exact success probability under coherent error is smaller than the results using Pauli twirling approximation at physical level. 
\end{abstract}


\maketitle

\section{Introduction}\label{Intro}

It is believed that quantum computers can solve useful problem much faster than the most powerful classical supercomputers~\cite{Nielsen&Chuang}. However, to realize meaningful calculations, one may need to build a large-scale quantum processor with noise mitigated or corrected to an extremely low level
Quantum error correction schemes~\cite{shor} was invented for more than twenty years and was then quickly developed in theory for encoding quantum information with a reduced error rate, and the famous examples include the $9$-qubit Shor code~\cite{shor}, the $7$-qubit Steane code~\cite{stean}, and the most promising surface code~\cite{kitaev,dennis,fowler}. 
The concept of fault-tolerant quantum computation~\cite{ShorIEEE1996} are introduced to ensure the quantum computer to operate at arbitrary logical accuracy for noisy devices, as long as the physical error rate is below an error threshold~\cite{Aharonov-threshold,Preskill1998,Knill1998}. There are two major types of fault-tolerance architecture: the concatenation architectures with recursive simulation~\cite{AGP} and the topological architectures~\cite{kitaev,dennis,fowler}. Due do the locality property of the physical device structure and high error threshold~\cite{dennis}, the surface code~\cite{kitaev,fowler} based on topological architecture is thought to be the most promising candidate for realizing fault-tolerant quantum computation. 

Recently, with the experimental development of quantum hardware~\cite{arute,Arute-Google2020Science,Gong-USTC2021-Science,Pino-TrapIon2021-Science,wu2021strong,ryananderson2021,zuchongzhiPRL,ZHU2021}, studying the performance of quantum error correction codes (QECCs) experimentally becomes a very important and hot topic~\cite{linke,andersen,mcewen,googleai}. However, the fault-tolerant encoding of a logical qubit and the full fault-tolerant QEC have never been demonstrated comprehensively. In fact, the theory of fault-tolerant quantum computation and QEC is only well-developed base on some key assumptions.
One of them is that all the noisy parts of the quantum computer are formulated for ideal stochastic error models. However, under more realistic situations, e.g. due to unavoidable imperfect qubit frequency, control pulse and measurement calibrations, the coherent errors on data qubits~\cite{Flammia-PRA,beale,Bravyi-NPJ,Iverson_2020} and detection-induced coherent errors~\cite{yang2021quantum} become important in quantum computer. 

The consequence from more general noise in quantum computation was recognized very early in the literature~\cite{Rahn-PRA-2002,aliferis,FernIEEE-2006}. In order to understand the fault-tolerance threshold from realistic coherent noise, recent studies considered the honest approximations~\cite{Puzzuoli-PRA-2014}, full-scale simulations for small codes~\cite{GutierrezPRA16}, tensor network representation for surface code~\cite{Darmawan-PRL-2017}. A numerical simulation~\cite{Bravyi-NPJ} based on Majorana fermion representation was carried out to study the toric code, for code distance up to $d=37$. Their numerical results indicate that the coherent error model may also has an error threshold, which is close to the threshold of the Pauli twirled stochastic error model~\cite{Bravyi-NPJ}.  However, more rigorous analytic investigations are necessary for understanding the structure of QEC and fault-tolerance under coherent errors. One of the key early analytic understandings is from the relation between the average gate infidelity and diamond distance~\cite{Beigi_2011,Wallman_2014,Sanders_2015,Kueng-PRL-2016}. The average gate infidelity $r(\mathcal{D})$ due to a noise process $\mathcal{D}$ can be accessed experimentally via randomized benchmarking~\cite{Emerson_2005,Emerson-Science-2007,Knill-RB-PRB-2008,Dankert-PRA-2009,Magesan-RB-PRL-2011}. While on the other hand, the error thresholds are usually obtain using rigorous bounds via the diamond distance  $D_{\diamond}(\mathcal{D})$~\cite{Kitaev_1997_QEC}, which cannot be obtained experimentally without complete process tomography. The diamond distance scales as $D_{\diamond}(\mathcal{D})\propto r(\mathcal{D})$ for stochastic Pauli noise, but scales differently as $D_{\diamond}(\mathcal{D})\propto \sqrt{r(\mathcal{D})}$ for the coherent noise~\cite{Kueng-PRL-2016}. Those results indicate that if the error threshold exists for the coherent error models, the threshold might be to square the values of their stochastic counterparts. Although the mitigation schemes~\cite{Wallman-RC-PRA-2016,Chamberland-PRA-2017,Chamberland2018faulttolerant,Debroy-PRL-2018,cai-npj-2020,Ouyang-npj-2021} are proposed for the coherent noise, the assumptions beyond the realistic situations are necessary for those schemes. 

Some recent analytic work has made great progress in understanding one type of coherent error models, e.g. the independent coherent noise~\cite{Flammia-PRA,beale,Iverson_2020}. Ref.~\cite{Flammia-PRA} proved an interesting result that the diamond distance $D^{L}_{\diamond}$ of the logical qubit is related to the the diamond distance of the physical qubit or the rotation angle $\theta$, i.e.  $D^{L}_{\diamond}\leq c_{n,k}|\sin\theta|^d$ for $[n,k,d]$ code, where $d$ is the code distance and the prefactor $c_{n,k}$ is upper bounded by an exponential function of the number of the encoded qubit $n$ (note that $n\sim 2d^2$ for surface code). Therefore, the results cannot lead to positive impression in the large code limit. In addition, even with a better prefactor, we notice that the logical diamond distance $D^{L}_{\diamond}\sim r^{(d+1)/2}$ for the stochastic noise and the average infidelity from the randomized benchmarking~\cite{Kueng-PRL-2016} could scale as $r\sim \theta^2$; and therefore, we have to be careful to reach the conclusion that the coherent errors are more favourable than the stochastic ones. Ref.~\cite{beale} also studied the independent coherent noise for stabilizer code and focused on the off-diagonal parts due to the noise channel in the logical code space. They found that after perfect syndrome measurement, the syndrome averaged logical off-diagonal terms of the error channel decays exponentially with respect to the code distance, and the decay is faster than that from the logical diagonal terms. Then, they made a conclusion that the syndrome measurements of the stabilizer code decohere independent coherent errors. Ref.~\cite{Iverson_2020} 
conducted a more comprehensive study for the independent coherent noise for the toric code. They went beyond the small $\theta$ limit~\cite{beale} to considered all corrections of $\theta$, and show that the logical channel after error correction loses coherence as increasing distance, which is consistent to the results from Ref.~\cite{beale}. However, those results mainly focused on the matrix elements between different logical states, i.e. logical off-diagonal terms. A natural question to ask is that how dose the coherency of the noise channel affect the logical diagonal terms and the success probability of error corrections.


\begin{center}
\textbf{{Summary of results}}
\end{center}

In this paper, we study the independent coherent error as well, and focus on topological surface code architecture. We aim to study the effects of the coherent error channel on the success probability of the error correction procedure. 
In order to develop our analytic approach, we first revisit the basic concept and the notations for the topological surface code~\cite{dennis,Bombin-arxiv} in the beginning of Sec.\ref{sec:procedure}. Then, we introduce the error correction procedure for the coherent error channel, and consider a maximum likelihood decoder to define the success probability of the error correction. 

In Sec. \ref{sec:distinguish}, we rewrite the coherent error after a perfect syndrome measurement into a quantum channel, and split its effect into two categories using the local operators on surface code: logical diagonal terms and logical off-diagonal terms. Ref.~\cite{beale} shows that the logical off-diagonal terms for stabilizer codes decay exponentially as increasing code distance, and thus conclude that QEC procedures decohere noise. However, we realized that coherent error may play an important role for the logical diagonal terms, and divide those parts into the physical diagonal terms and the physical off-diagonal terms, which correspond to the stochastic contribution with Pauli twirl approximation and coherent contribution beyond Pauli twirling, respectively.

In Sec. \ref{sec:KL}, we study the coherent error channel in the framework of the Knill-Laflamme (K-L) criterion or the QEC condition~\cite{KL-PRA97}, and find that the surface code under the coherent error channel satisfies a generalized K-L criterion~\cite{beny2010general} and falls into the category of the approximate QEC. We find the extra term $\epsilon$, that roughly describing the accuracy of the approximate QEC, corresponds to the coherent part of the error channel at the logical level; and therefore, for a fixed small coherent rotation angle, the coherent error channel converges towards probabilistic quantum channel as code distance approaches infinity. On the other hand, we also find that if the code with a fixed distance $d$ is $\epsilon$- correctable, the value of $\epsilon$ describing the accuracy of the approximate QEC cannot be smaller than a lower bound. 

In Sec. \ref{sec:compare}, we focus on the success probability of the QEC procedure by considering the logical diagonal terms, and aim to compare the effects of coherent parts with the stochastic parts (terms after Pauli twirling). One can imagine those stochastic parts as the Pauli averaged infidelity of the coherent errors. We analytically find that the exact success probability from the contribution of the coherent parts is always smaller than the contribution from the stochastic parts using Pauli twirling at physical level, and our analytic results are consistent with the numerical simulations~\cite{Bravyi-NPJ}. 
We can see that even if the logical off-diagonal terms due to the coherent errors decay as increasing the code distance, the success probability determined by the logical diagonal terms will show distinct behaviors regarding the coherency of the error channel. 
The conclusions and discussions are shown in Sec. \ref{sec:con}.


\section{Preliminary of QEC in surface code}\label{sec:procedure}

We consider the quantum error correction procedure based on the topological surface code~\cite{kitaev,dennis}, and focus on their performance under the coherent errors, e.g. coherent rotations due to imperfect control pulse calibrations. We assume the coherent noise process is independently applied on each qubit in the code, and the coherent rotations for all the qubit are along the $X$ axis with the same rotation angle $\theta$.

\begin{figure}
	\centering
	\includegraphics[scale=0.3]{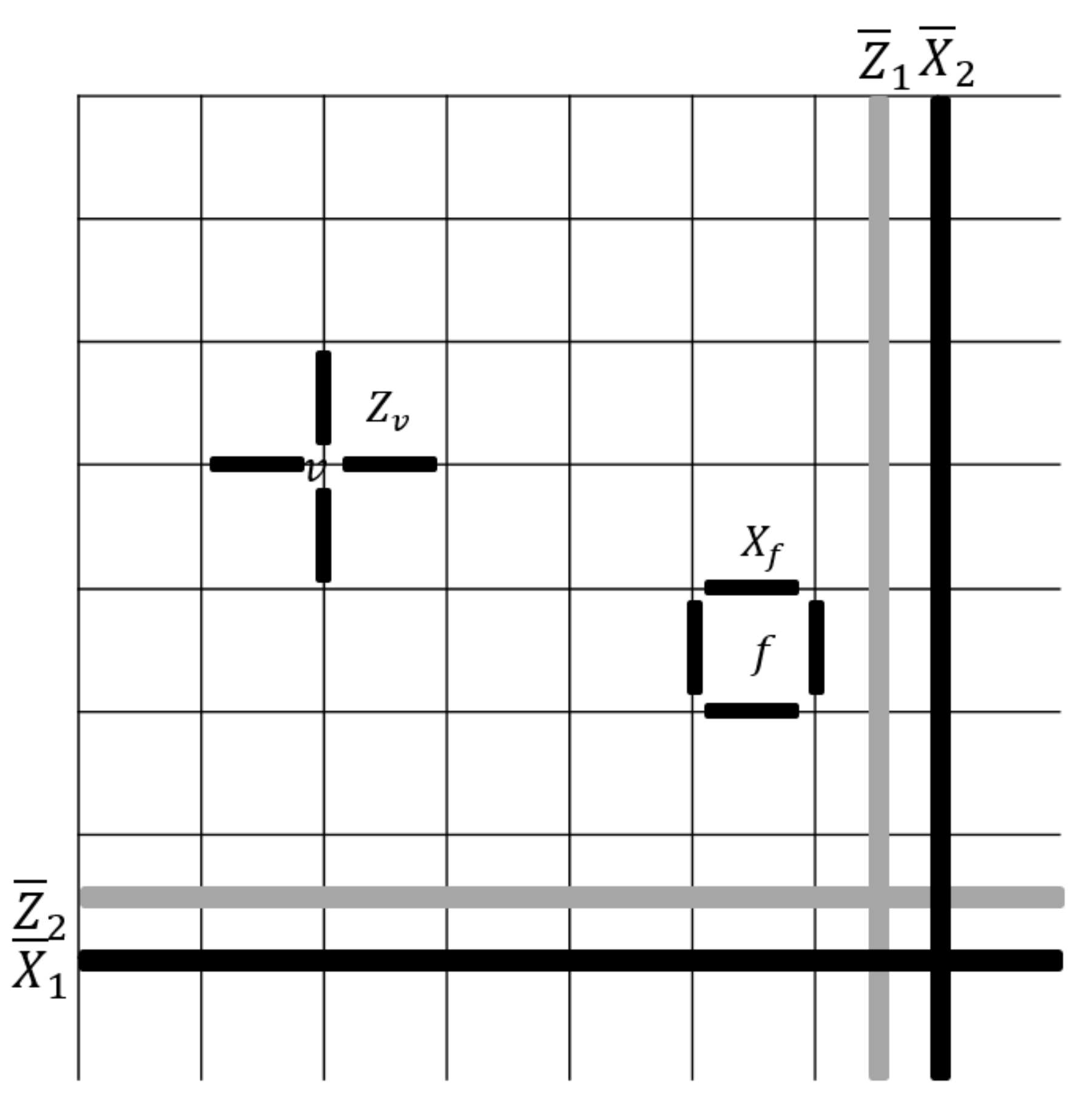}
	\caption{Stabilizer generators and logical Pauli operators of toric code (the lattice has periodic boundary condition).}
	\label{fig:generator}
\end{figure}

Let us briefly review the QEC procedure of the surface code on a torus, which is also called toric code \cite{kitaev,dennis}, and define the notation used in the work. Toric code is a stabilier code defined on a two dimensional square lattice with periodic boundary condition, like in FIG. \ref{fig:generator}. The qubits lies on the edges of the lattice. There are two kinds of stabilizer generators $Z_v$ and $X_f$. $Z_v$ means four Pauli $Z$ operators act around a vertex $v$, and $X_f$ means four Pauli $X$ operators act around a face $f$, as in FIG. \ref{fig:generator}. There are $4$ logical Pauli operators $\overline{X}_1$, $\overline{X}_2$, $\overline{Z}_1$, $\overline{Z}_2$. The logical $X$ operators are non-contractable loops on the lattice, and the logical $Z$ oerators are non-contractable loops on the dual lattice. The code subspace is $4$ dimensional ($2$ logical qubits), and the basis states can be chosen as eigenstates of $\overline{Z}_1$ and $\overline{Z}_2$, denoted as $\ket{0,0}$, $\ket{0,1}$, $\ket{1,0}$, $\ket{1,1}$. Logical Pauli operators act on these logical qubits just as physical Pauli operators act on Physical qubits.

\begin{figure}[t]
	\centering
	\includegraphics[scale=0.3]{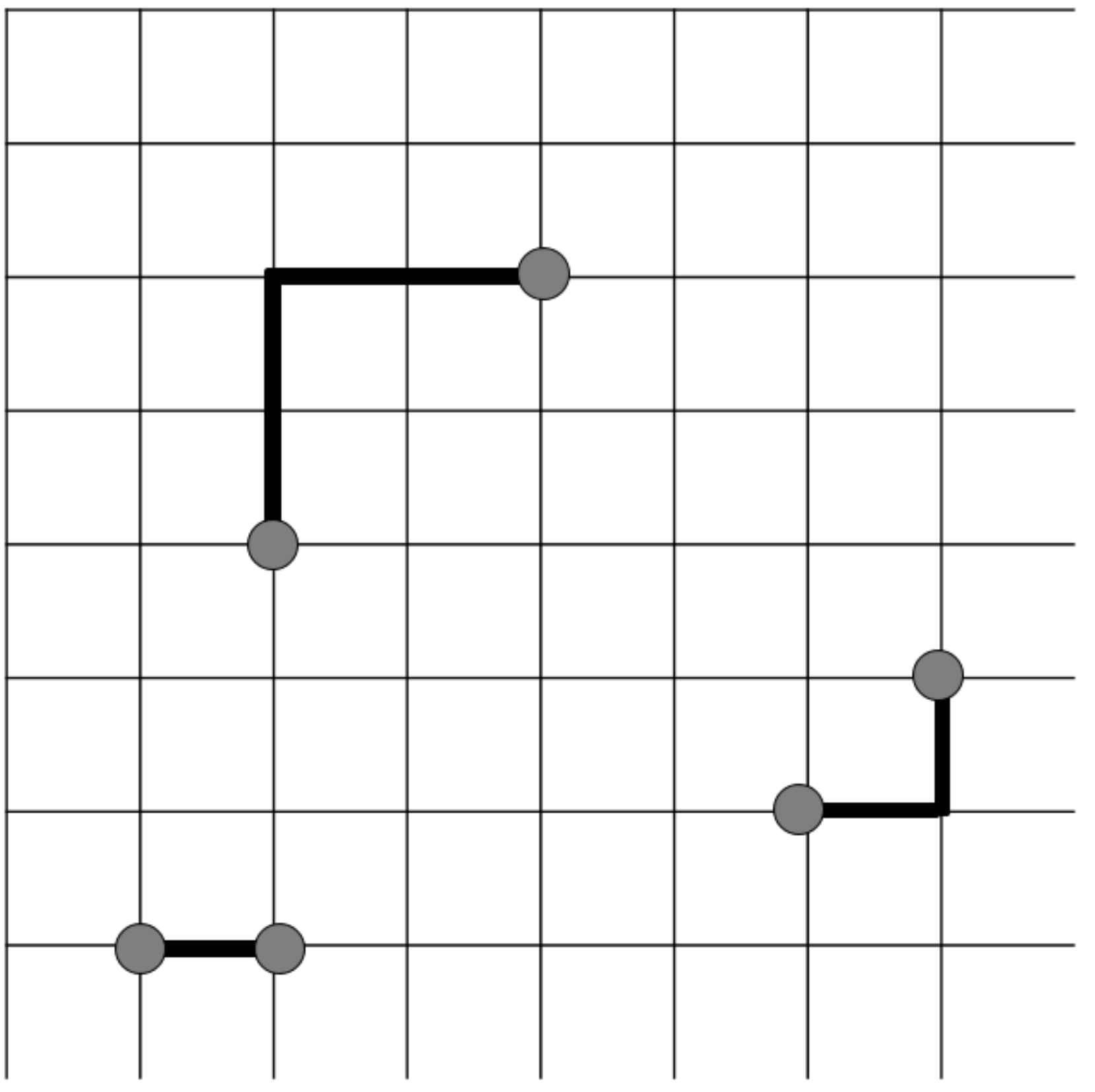}
	\caption{Error chain on the toric code lattice.}
	\label{fig:errorchain}
\end{figure}

\begin{figure}
	\centering
	\includegraphics[scale=0.3]{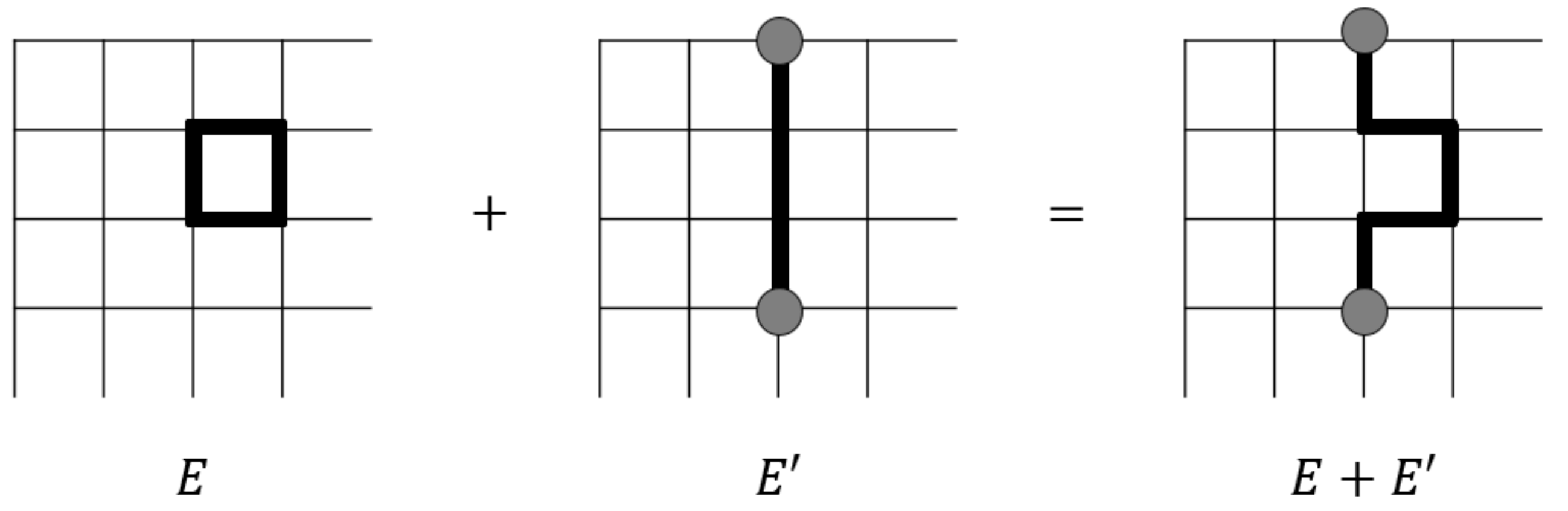}
	\caption{Error chain addition.}
	\label{fig:addition}
\end{figure}

When Pauli $X$ errors occurs on Physical qubits, they can be viewed as error chain on the toric code lattice, as in FIG. \ref{fig:errorchain}. Since Pauli $Z$ errors can be dealt with similarly on the dual lattice, we consider only Pauli $X$ errors. In order to write down these error chains explicitly, we follow the notation in Ref. \cite{Bombin-arxiv}. Suppose the number of vertices, edges and faces are respectively $\mathcal{V}$, $\mathcal{E}$ and $\mathcal{F}$.  We assign a $\mathbb{Z}_2$ group on each edge. The tensor product of these $\mathbb{Z}_2$ forms an Abelian group $C_1 \simeq \mathbb{Z}_2^\mathcal{E}$. Any element $E \in C_1$ is a binary vector of length $\mathcal{E}$, and can be expressed as
\begin{equation}
    E=\sum_i c_i e_i, \quad c_i = 0 \text{ or } 1,
    \label{eq:ErrorChain}
\end{equation}
where $\{e_i\}_i$ stands for the edges. $E$ can also be viewed as an error chain on the lattice. For example, the error chain in FIG. \ref{fig:errorchain} will be expressed as $E=\sum_i c_i e_i$ where $c_i=1$ for all the bold edges and $c_i=0$ for all the empty edges. The corresponding Pauli operator will be denoted as $X_E$. The binary addition of two error chains $E=\sum_i c_i e_i$ and $E' = \sum_i c'_i e_i$ is defined as $E+E'= \sum_i (c_i+c'_i)_{\mathrm{mod} 2} e_i$, visually expressed as FIG. \ref{fig:addition}. 
Obviously we have
\begin{equation}
    X_{E} X_{E'} = X_{E+E'}
\end{equation}
for arbitrary $E, E' \in C_1$.
We can define Abelian groups for vertices and faces in similar ways, denoted as $C_0$ and $C_2$. Naturally there are boundary operators between $C_2$, $C_1$ and $C_0$:
\begin{equation}
    C_2 \stackrel{\partial_2}{\longrightarrow} C_1 \stackrel{\partial_1}{\longrightarrow} C_0.
    \label{eq:complex}
\end{equation}
\begin{figure}
	\centering
	\includegraphics[scale=0.3]{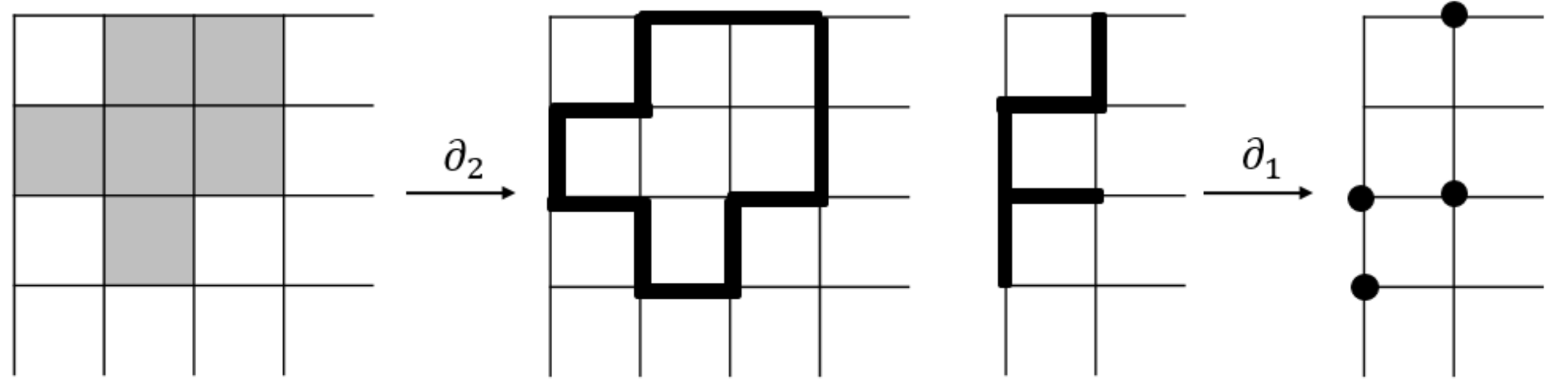}
	\caption{Boundary operators.}
	\label{fig:boundary}
\end{figure}
The action of boundary operators is shown in FIG. \ref{fig:boundary}. Any error chain that represents a $X$ stabilizer is boundary of a region, so it belongs to the image of boundary operator on faces $B_1=\Im \partial_2$. Two error chains that differ by a stabilizer (differ by a boundary) have the same effect on code subspace. We use $[\cdot]$ to denote the equivalent class module $B_1$, $[E]= \{E+b|b\in B_1\} \in C_1/B_1$. $C_1/B_1$ describes the effect of error chains on cod subspace. Clearly $[E]+[E']=[E+E']$. Suppose error $X_E$ occurs, if we perform syndrome measurement of all $Z_v$ operators, we can detect all the ends of error chain $E$, which is $S=\partial_1 E$. $S$ is what we called syndrome. Error correct introduces a new Pauli operator $X_{E_c}$ that form closed loop configuration with original error. Closed loop configurations are just error chains that have no ends, so they belong to the kernel of boundary operator on edges $Z_1=\Ker \partial_1$. The boundary of regions must be closed loops, so $B_1 \subset Z_1$, in other word $\partial^2=\partial_1 \circ \partial_2=0$.  A closed loop configuration might purely be a stabilizer, or it might contain non-contractable loops, which correspond to logical operators. Consider the quotient $H_1 = Z_1/B_1$. $H_1$ has four elements on torus. For $[L]\in H_1$, if $L$ is empty error chain $L=0$, $[0]$ is just the set of boundaries, the set of stabilizers. if $[L] \neq [0]$, $X_L$ is equivalent to one of the three logical operators $\overline{X}_1,\overline{X}_2,\overline{X}_1\overline{X}_2$. In the error correction procedure, if the final loop configuration $E+E_c$ belongs to $[0]$, the error correction succeed. But if $E+E_c \in [L\neq 0]$, there will be a logical error.

\subsection{Error Correction Procedure for surface code}\label{subsec:procedure}

The error correction procedure for surface code with coherent errors can be summarized as the following four steps:
\begin{enumerate}
	\item Start with the initial state $\ket{\psi_0}$ which belongs to the code space of the surface code,
	\item Coherent error $U=\prod_e e^{\imi \theta X_e}$ occurs. ($X_e$ stands for Pauli $X$ operator acting on qubit $e$),
	\item Apply perfect syndrome measurements for all stabilizer operators,
	\item Apply Pauli correction $X_{E_c}$ depending on the decoder and the syndrome.
\end{enumerate}

We expand the coherent error operator $U$ in order of the Pauli $X$ operators:
\begin{equation}
		U =  \prod_e (\cos \theta + \imi \sin \theta X_e)
		= \sum_{E} a_E X_E.
	\label{eq:error}
\end{equation}
Here $\sum_E$ means the summation over all error chains, and the coefficient can be written as
\begin{equation}
	a_E = \prod_{e\in E} \imi\sin\theta \prod_{e \notin E} \cos \theta.
\end{equation}
It shows that the coherent error cause a superposition of different error chain operators. Applying this operator on the initial state leads to a coherent liner superposition of error chain states  
\begin{equation}
\ket{\psi_{CE}}=\sum_E a_E X_E \ket{\psi_0}.
\end{equation}
After syndrome measurement for all the stabilizer operators, we get an ensemble of the possible syndrome outcomes:
\begin{equation}
	\{ \quad N_S\sum_{E|\partial E=S} a_E X_E \ket{\psi_0}  \quad \}_S,
	\label{syndrome}
\end{equation}
where $S$ denotes the syndrome for a particular measurement outcome, $N_S$ is the normalization factor, $\sum_{E|\partial E =S}$ means the summation over all error chains that have syndrome $S$. The ensemble is weighted by the probability of different
syndrome outcomes:
\begin{equation}
	\mathrm{Pr}(S) = \sum_{E| \partial E = S}\sum_{E'|\partial E' = S} a_E a_{E'}^* \bra{\psi_0} X_E X_{E'} \ket{\psi_0}.
	\label{eq:syndrome-probability}
\end{equation}
The normalization factor $N_S$ equals to $1/\sqrt{\mathrm{Pr}(S)}$.
If we readout the measurement outcome, we will get a specific syndrome $S$.
Then after Pauli correction, the final state is:
\begin{equation}
	\ket{\psi_{final}} = N_S\sum_{E|\partial E=S} a_E X_{E+E_c} \ket{\psi_0}.
\end{equation}

We want to understand the key difference between this result and that using Pauli twirl approximation~\cite{Silva-PRA}, where the coherent error channel on each qubit $\Lambda(\rho) = \e{{\imi \theta X}}\rho \e{{-\imi \theta X}}$ is averaging over a Pauli twirl $\overline{\Lambda} (\rho) = \cos^2\theta \rho + \sin^2\theta X\rho X$. In that case, we instead obtain the same ensemble 
as those from the stochastic error model:
\begin{equation}
	\{\quad a_E X_E \ket{\psi_0} \quad\}_E,
	\label{ensemble-E}
\end{equation}
which is weight by the probability of different errors $p(E) = |a_E|^2$. Note that the state from the ensemble due to the stochastic error can be simply described by error chain applying onto the code space state. On the other hand, the coherent error channel generate a coherent linear superposition of error chain states.

\subsection{Success Probability of Error Correction}\label{subsec:probability}
To characterize the effectiveness of error correction, we use the fidelity $F=|\braket{\psi_0|\psi_{final}}|^2$ to represent how close are the final state and the initial state, and $F$ is close to $1$ when the final state gets close to the initial state.
We can also view the fidelity $F$ as the success probability of error correction.
Note that this expectation value depends on the choice of initial state. If the initial state 
is an eigenstate of logical Pauli $X$ operator, then the final state differs from the initial state by a global 
phase factor, which means the error correction will always succeed. But if the initial state is 
an eigenstate of logical Pauli $Z$ operator, the Pauli $X$ error will have the most significant influence on error correction.
So in the following sections, we assume the initial state is the eigenstate of both logical $Z$ operators for the toric code, labeled by $\ket{\psi_0}=\ket{0,0}$ (recall that the four eigenstates of the two logical $Z$ operators are $\ket{\psi_0}=\ket{0,0}$, $\ket{\psi_0}=\ket{0,1}$, $\ket{\psi_0}=\ket{1,0}$, and $\ket{\psi_0}=\ket{1,1}$). So we have:
\begin{equation}
    \begin{aligned}
    	&|\braket{\psi_0|\psi_{final}}|^2 = |N_S\sum_{E|\partial E=S} a_E  \bra{0,0} X_{E+E_c} \ket{0,0}|^2\\
    	&= |N_S\sum_{E\in[E_c]} a_E|^2.
    \end{aligned}
	\label{eq:expect}
\end{equation}
Here the second line is due to the fact that $\bra{0,0}X_{E+E_c}\ket{0,0}$ equals $1$ if $E$ is in the same equivalent class of $E_c$ which means $X_{E+E_c}$ is a stabilizer and it vanishes if $E+E_c$ contains non-contractable loops which means $X_{E+E_c}$ is a logical operator on code subspace. Here we note that the dependence of fidelity on initial state is not a consequence of coherence. For example for stochastic Pauli $X$ error, if the initial state is a eigenstate of logical $X$ operators, then after error correction will always be the initial state times a overall phase factor, and the fidelity is always $1$.

For convenience when comparing coherent error with stochastic error, we can explain the above expression of fidelity as probability. We note that the state after syndrome measurement is $N_S\sum_{E|\partial E=S} a_E X_E \ket{0,0}$.
Since a specific syndrome corresponds to four equivalent classes of error chain, 
we can split the summation $\sum_{E|\partial E=S}$ to the summation of four different error classes:
\begin{equation}
	\begin{aligned}
		&N_S\sum_{E|\partial E=S} a_E X_E \ket{0,0} =\\
		&  (N_S\sum_{E\in[E_{S,1}]} a_E) X_{E_{S,1}} \ket{0,0} + (N_S\sum_{E\in[E_{S,2}]} a_E) X_{E_{S,2}} \ket{0,0}\\
		& + (N_S\sum_{E\in[E_{S,3}]} a_E) X_{E_{S,3}} \ket{0,0} + (N_S\sum_{E\in[E_{S,4}]} a_E) X_{E_{S,4}} \ket{0,0}.
	\end{aligned}
\end{equation}
where $[E_{S,1}]$, $[E_{S,2}]$, $[E_{S,3}]$, $[E_{S,4}]$ are the four error classes. The error chain $E_{S,i}$ is labeled with $S$ to notify that the syndrome of $E_{S,i}$ is $S$, $\partial E_{S,i}=S$. Therefore, the probability of error class $[E_{S,i}]$ under the condition of syndrome $S$ yields
\begin{equation}
  \mathrm{Pr}([E_{S,i}]|S)=|N_S\sum_{E\in[E_{S,i}]} a_E|^2.
\end{equation}
Besides, the probability is normalized:
\begin{equation}
	\begin{aligned}
		&\sum_{[E_{S,i}]|\partial E_{S,i} =S} \mathrm{Pr}([E_{S,i}]|S) = 1,\\
		&\sum_S \sum_{[E_{S,i}]|\partial E_{S,i} =S} \mathrm{Pr}(S) \mathrm{Pr}([E_{S,i}]|S) = 1,
	\end{aligned}
\end{equation}
Here $\sum_{[E_{S,i}]|\partial E_{S,i} =S}$ means the summation over all four error classes that have syndrome $S$.
because the states we considered are all normalized.
So we call $\mathrm{Pr}([E_c]|S)$ the success probability of the error correction procedure, and it is just the fidelity $F$ we concerned.

With $\mathrm{Pr}([E_{S,i}]|S)$, we can define a maximum likelihood decoder \cite{Flammia-arxiv}, which is to choose the error class $[E_{S,max}]$ that has the largest $\mathrm{Pr}([E_{S,i}]|S)$ for Pauli correction.
In that case, the logical class in the quantum channel from initial state to final state $\sum_{E|\partial E=S} a_E X_{E+E_{S,max}}$ is $[0]$, which corresponds to identity operator,
so the final state is most close to the initial state.

Here we introduce some other notations that will be used later.
For an error class $[E_{S,i}]$ under a given syndrome $S$, we can always split it as the summation of $[E_{S,max}]$ and 
a logical class $[L]$, $[E_{S,i}] = [E_{S,max}]+[L]$. So we can equivalently denote $\Pr([E_{S,i}]|S) = \Pr([E_{S,max} +L] |S)$ as $\Pr([L]|S)$. The product of it and syndrome probability $\Pr(S)$ is the joint probability
\begin{equation}
    \mathrm{Pr}([L],S) = \Pr(S)\Pr([L]|S) = |\sum_{E\in[E_{S,max}+L]} a_E|^2,
    \label{eq:joint-prob}
\end{equation}
and obviously $\sum_{[L],S} \mathrm{Pr}([L],S) = 1$, $\sum_{[L]}\mathrm{Pr}([L],S) = \mathrm{Pr}(S)$.
In this case $[L]$ and $S$ can be viewed as two random variables and now the choice of the value of $[L]$ do not rely on $S$ while $[E_{S,i}]$ must have $S$ as its ends. Then $\Pr([L=0]|S)$ is the success probability of the decoder for a particular syndrome $S$, and $\sum_S \Pr([0],S)$ is the averaged success probability of the decoder.

\section{Distinguish Diagonal and Off-diagonal Parts}\label{sec:distinguish}
In this section we distinguish coherent contribution and stochastic contribution in the error channel. Coherent error an stochastic error can be related through Pauli twirling. The single qubit coherent error channel we considered is:
\begin{equation}
	\begin{aligned}
		&\Lambda(\rho) = \e{{\imi \theta X}}\rho \e{{-\imi \theta X}}  \\
		&= \cos^2\theta I\rho I + \imi \sin\theta \cos\theta X \rho I\\
		& - \imi \sin\theta \cos\theta I\rho X + \sin^2\theta X \rho X.
	\end{aligned}
\end{equation}
After Pauli twirl approximation, which ignores the off-diagonal part of error channel $\Lambda$, we get the Stochastic error channel:
\begin{equation}
	\overline{\Lambda} (\rho) = \cos^2\theta \rho + \sin^2\theta X\rho X.
\end{equation}
$\overline{\Lambda}$ is the error channel that Pauli $X$ error occurs at probability $p=\sin^2 \theta$. Generally we can say that stochastic contribution is the diagonal part of the error channel.

We know that the state after syndrome measurement is $N_S \sum_{E|\partial E = S} a_E X_E \ket{\psi_0}$.
The overall error channel is:
\begin{equation}
    \begin{aligned}
        &\mathcal{D}(\ket{\psi_0}\bra{\psi_0})=\\
        &\sum_S \sum_{E|\partial E = S}\sum_{E'|\partial E' = S} a_E a_{E'}^* X_E \ket{\psi_0}\bra{\psi_0}X_{E'}.
    \end{aligned}
\end{equation}
After we gather the information of syndrome, the error channel of a particular syndrome $S$ is
\begin{equation}
    \begin{aligned}
        &\mathcal{D}_S(\ket{\psi_0}\bra{\psi_0})=\\
        &|N_S|^2 \sum_{E|\partial E = S}\sum_{E'|\partial E' = S} a_E a_{E'}^* X_E \ket{\psi_0}\bra{\psi_0}X_{E'}.
    \end{aligned}
	\label{eq:dm}
\end{equation}
At the level of physical noise, the diagonal (off-diagonal) terms of coherent error are $E=E'$ ($E\neq E'$) terms in \eqref{eq:dm}, we will call them physical diagonal (off-diagonal) terms.
Pauli twirl approximation means that all the physical off-diagonal terms are dropped.
In \eqref{eq:dm}, the summation of error chain $\sum_{E|\partial E =S}$ can be split to summation of logical classes $\sum_{[L]}$ and 
summation of errors in a logical class $\sum_{E_L \in [E_{S,max} + L]}$:
\begin{equation}
    \begin{aligned}
        &\mathcal{D}_S(\ket{\psi_0}\bra{\psi_0})\\
    	&=|N_S|^2 \sum_{\substack{[L] \\ [L']}} \sum_{\substack{E_L \in [E_{S,max} + L] \\ E'_{L'} \in [E_{S,max} + L']}} a_{E_L} a_{E'_{L'}}^* X_{E_L} {\ket{\psi_0}} {\bra{\psi_0}} X_{E'_{L'}}\\
    	&=|N_S|^2 \sum_{\substack{[L] \\ [L']}} \left(\sum_{\substack{E_L \in [E_{S,max} + L] \\ E'_{L'} \in [E_{S,max} + L']}} a_{E_L} a_{E'_{L'}}^*\right) X_{L} X_{E_{S,max}}\\ &\qquad\qquad\qquad\qquad\qquad\qquad \times\ket{\psi_0}\bra{\psi_0}X_{E_{S,max}} X_{L'}.
    \end{aligned}
    \label{eq:density-matrix}
\end{equation}
Here $\sum_{[L]}$ means the summation over all logical class in $H_1$, and the third line is because operators that belong to the the equivalent class has the same effect on code subspace.
In \eqref{eq:density-matrix}, the terms for $[L]=[L']$ ($[L] \neq [L']$) correspond to the diagonal (off-diagonal) parts of the matrix elements in the logical code space, we call them logical diagonal terms.
If we apply Pauli correction according to the decoder, the error chain decoder chooses $[E_{S,max}]$, then the final error channel becomes
\begin{equation}
    \begin{aligned}
        &\mathcal{N}_S(\ket{\psi_0}\bra{\psi_0})=\\
        &|N_S|^2 \sum_{\substack{[L] \\ [L']}} \left(\sum_{\substack{E_L \in [E_{S,max} + L] \\ E'_{L'} \in [E_{S,max} + L']}} a_{E_L} a_{E'_{L'}}^*\right) X_{L}\ket{\psi_0}\bra{\psi_0}X_{L'}.
    \end{aligned}
    \label{eq:overall-channel}
\end{equation}
Then, we can see that when choosing $\ket{\psi_0}$ to be the eigenstate of logical $Z$ operator, only logical diagonal terms contribute to the fidelity $F = |\braket{\psi_0|\psi_{final}}|^2= \bra{\psi_0} \mathcal{N}_S(\ket{\psi_0}\bra{\psi_0}) \ket{\psi_0}$ or the success probability, which can be written as
\begin{equation}
    \Pr([E_{S,max}]|S)=F=|N_S|^2\sum_{E,E' \in [E_{S,max}]}  a_{E} a_{E'}^*.
\end{equation}
Similarly, for other logical class,
\begin{equation}
    \Pr([E_{S,max}+L]|S)=|N_S|^2\sum_{E_L,E'_L \in [E_{S,max}+L]} a_{E_L} a_{E'_{L}}^*,
    \label{eq:sucprob}
\end{equation}
which describes the probability for the case with logical errors $L$. Therefore, $\Pr([E]|S)$ can also be viewed as the coefficients of logical diagonal terms.

To characterize the influence of Pauli twirl approximation, we need to compare the contribution of coherent (off-diagonal) part with the contribution of stochastic (diagonal) part in $\mathrm{Pr}([E]|S)$.
In \eqref{eq:density-matrix}, we clearly see that the physical diagonal terms are terms with $E_L=E'_L$ and the physical off-diagonal terms are those with $E_L \neq E'_L$. So in \eqref{eq:sucprob}, the stochastic contributions are the $E_L=E'_L$ terms in the summation and the coherent contributions are the $E_L \neq E'_L$ terms in the summation.

Reference \cite{beale} proved that logical off-diagonal terms decay exponentially with respect to code distance, and we will discuss the application of the result~\cite{beale} for the surface code system in appendix \ref{app:logical}. Although the logical off-diagonal terms are proven to decay, there is no discussion about the coherent error effect due to the physical off-diagonal parts as defined below Eq.\eqref{eq:dm}, which are part off the "logical diagonal" terms. That means coherence at the level of physical error chain could still affect the result of error correction and makes Pauli twirl approximation inaccurate. We will discuss this topic in section \ref{sec:compare}.

\section{Knill-Laflamme Criterion}\label{sec:KL}
We can analyse coherent error with Knill-Laflamme (K-L) criterion for quantum error correction code~\cite{KL-PRA97,knill}.
K-L criterion states that~\cite{KL-PRA97}: for an error process
\begin{equation}
	\mathcal{D}(\rho) = \sum_{K_a \in S_D} K_a \rho K_a^\dagger,\quad \sum_{K_a \in S_D} K_a K_a^\dagger = I
\end{equation}
acting on quantum code $C$, a subset of errors $S_\epsilon \subset S_D$ can be recovered if and only if for arbitrary basis states $\ket{i},\ket{j}\in C$ and $K_a,K_b \in S_\epsilon$,
\begin{equation}
	\braket{i|K_a^\dagger K_b|j} = \lambda_{ab}\delta_{ij}.
	\label{eq:KL}
\end{equation}
$\lambda_{ab}$ is an arbitrary Hermitian matrix, and does not depend on $i,j$.
For the coherent error, we consider the quantum channel that coherent rotation shown in Eq. \eqref{eq:error}, and then perform the perfect syndrome measurement. In this case, we obtain an equivalent error channel $\mathcal{D}(\rho)$ described by an ensemble shown in Eq. \eqref{syndrome}, and the corresponding quantum operator representation can be written as
\begin{equation}
		\mathcal{D}(\rho) = \sum \mathrm{Pr}(S) K_S \rho K_S^\dagger
\end{equation}
where the Kraus operators are
\begin{equation}
  K_S = N_S\sum_{E|\partial E=S} a_E X_{E}
\end{equation}
and the syndrome probability $\mathrm{Pr}(S)=1/|N_S|^2$ is defined in \eqref{eq:syndrome-probability}. It is easy to verify the completeness of the Kraus operators:
\begin{align}
    &I = U U^\dagger = \sum_{E} a_E X_{E} \sum_{E'} a_{E'}^* X_{E'}  \nonumber\\
    =&\sum_{S} \sum_{E|\partial E=S} a_E X_{E}\sum_{S'} \sum_{E'|\partial E'=S'} a_{E'}^* X_{E'}\nonumber\\
    =&\sum_{S} \sum_{E|\partial E=S} a_E X_{E} \sum_{E'|\partial E'=S} a_{E'}^* X_{E'}\nonumber\\
    =&\sum_{S} \mathrm{Pr}(S) K_S K_S^\dagger.
\end{align}
The reason the third line holds is that in $\sum_{E,E'}a_E a_{E'} X_{E}X_{E'}$ only the coefficient of identity is nonzero and the coefficient of other Pauli operators vanishes. That means the product of different syndrome vanish in the second line.

Let's examine if the coherent error model is consistent with the K-L criterion by evaluating the matrix elements $\braket{i|K_{S'}^\dagger K_S|j}$, $\ket{i},\ket{j} \in \{\ket{0,0},\ket{0,1},\ket{1,0},\ket{1,1}\}$. For the errors from different syndrome, i.e. $S\neq S'$, the K-L matrix elements must be zero because the states $\ket{i}$ and $\ket{j}$ in $\braket{i|K_{S'}^\dagger K_S|j}$ will be mapped to different syndrome and then become orthogonal. For the errors within the same syndrome $S=S'$, the K-L matrix elements are finite. However, since four logical classes are all contained in $K_S^\dagger K_S$, it is easy to show that the elements $\braket{i|K_S^\dagger K_S|j}$ are not necessary to be proportional to $\delta_{ij}$. Therefore, the coherent error model does not satisfy the standard K-L criterion.

However, we can show that coherent error satisfies the generalized K-L criterion~\cite{beny2010general}.
the generalized K-L criterion~\cite{beny2010general} states that:
for error process $\mathcal{N} (\rho) = \sum_k K_a \rho K_a^\dagger$ acting on quantum code $C$, The code is $\epsilon$-correctable
under $\mathcal{N}$ if and only if for arbitrary basis states $\ket{i},\ket{j}\in C$ and $K_a,K_b$:
\begin{equation}
	\braket{i|K_a^\dagger K_b|j} = \lambda_{ab}\delta_{ij} + \braket{i|B_{ab}|j},
	\label{eq:gKL}
\end{equation}
and $\Delta(\Lambda+\mathcal{B},\Lambda) \leq \epsilon$, where $\Lambda(\rho)=\sum_{ab}\lambda_{ab} \Tr(\rho) \ket{a}\bra{b}$ and $\mathcal{B}(\rho)= \sum_{ab} \Tr(\rho B_{ab}) \ket{a}\bra{b}$. In the above statements, $\Delta$ is called Bures distance and it measures the difference between two quantum channels. It is defined as
\begin{equation}
    \Delta(\mathcal{N},\mathcal{M}) = \sqrt{1-F(\mathcal{N},\mathcal{M})}.
\end{equation}
$F$ is the worst-case entanglement fidelity. It characterises the fidelity between two quantum channels and is defined as
\begin{equation}
	\begin{aligned}
		&F(\mathcal{N},\mathcal{M}) = \min_\rho F_\rho(\mathcal{N},\mathcal{M})\\
		& F_\rho(\mathcal{N},\mathcal{M}) = f(\mathcal{N}\otimes\mathrm{id} (\ket{\psi}\bra{\psi}),\mathcal{M}\otimes\mathrm{id} (\ket{\psi}\bra{\psi})).
	\end{aligned}
\end{equation}
Here $F_\rho(\mathcal{N},\mathcal{M})$ is called entanglement fidelity, $\ket{\psi}$ is a purification of $\rho$, $f$ is the fidelity between two states
\begin{equation}
    f(\rho,\sigma)=\Tr \sqrt{\sqrt{\rho}\sigma\sqrt{\rho}}.
\end{equation}
The definition of $\epsilon$-correctable is that there exists a recovery channel $\mathcal{R}$ such that:
\begin{equation}
	\Delta(\mathcal{RNC},I) \leq \epsilon.
	\label{eq:epsilon}
\end{equation}
Here $\mathcal{C}$ is the encode map that map the logical Hilbert space to the physical Hilbert space.
Note that $\mathcal{R}$ determined by K-L criterion might not be the Pauli correction we considered. Compared with normal K-L criterion, equation \eqref{eq:gKL} has an extra term $B_{ab}$. If the general K-L condition is satisfied, then the error is $\epsilon$-correctable, which tells us that the error might not be able to be completely corrected, but we can find a recovery channel so that the final channel can be close to identity.

We then consider our our error correction procedure for the coherent errors, and examine if the coherent errors can be described by the generalized K-L criterion~\cite{beny2010general}. Since the K-L matrix elements must be zero for errors from different syndrome, we only focus on the cases with the same syndrome below
\begin{widetext}
\begin{equation}
	\begin{aligned}
		\mathrm{Pr}(S) \braket{i|K^\dagger_S K_S|j} &= \sum_{E|\partial E=S}\sum_{E'|\partial E'=S} \braket{i|a^*_{E'} a_{E} X_{E'} X_{E} |j}\\
		&= (\sum_{[L]} \sum_{E_L,E'_L \in [E_{S,max} + L]} a^*_{E'} a_{E}) \delta_{ij} 
		+ \sum_{[L]\neq[L']} \sum_{E_L \in [E_{S,max} + L]} \sum_{E'_{L'} \in [E_{S,max} + L']} \braket{i|a_{E_L} a^*_{E'_{L'}} X_{E_L} X_{E'_{L'}}|j}\\
		&= (\sum_{[L]} \sum_{E_L,E'_L \in [E_{S,max} + L]} a^*_{E'} a_{E}) \delta_{ij}
	       + \sum_{[L]\neq[L']} \sum_{E_L \in [E_{S,max} + L]} \sum_{E'_{L'} \in [E_{S,max} + L']} \braket{i|a_{E_L} a^*_{E'_{L'}} X_{L+L'} |j}\\
		&= (\sum_{[L]} \sum_{E_L,E'_L \in [E_{S,max} + L]} a^*_{E'} a_{E}) \delta_{ij} + 
		\sum_{[L]} \braket{i|X_{L}|j} \sum_{[J] \neq [0]} \sum_{E_J \in [E_{S,max} + J]} \sum_{E_{J+L} \in [E_{S,max} + J+L]} a_{E_J} a^*_{E_{J+L}}.
	\end{aligned}
\end{equation}
So, we obtain the matrix of the generalized K-L criterion shown in Eq.(\ref{eq:gKL}):
\begin{equation}
	\begin{aligned}
		&\lambda_{S,S'} =\delta_{S,S'} \sum_{[L]} \sum_{E_L,E'_L \in [E_{S,max} + L]} a^*_{E'} a_{E} =\delta_{S,S'} \mathrm{Pr}(S) \overline{\mathcal{N}}(S)_{0,0}\\
		&B_{S,S'} = \delta_{S,S'} \sum_{[L] \neq [0]} X_{L} \sum_{[J]} \sum_{E_J \in [E_{S,max} + J]} \sum_{E_{J+L} \in [E_{S,max} + J+L]} a_{E_J} a^*_{E_{J+L}}=\delta_{S,S'} \mathrm{Pr}(S) \sum_{[L] \neq [0]} X_{L} \overline{\mathcal{N}}(S)_{0,L}.
	\end{aligned}
	\label{eq:lambdaB1}
\end{equation}
Here the quantity $\overline{\mathcal{N}}(S)_{L,L'}$ is expressed as:
\begin{equation}
    \overline{\mathcal{N}}(S)_{L,L'} 
	=\frac{1}{\Pr(S)}\sum_{[J]}\sum_{E \in [E_{S,max} + J + L]}  \sum_{E' \in [E_{S,max} + J + L']}a_{E} a_{E'}^*.
\end{equation}
This quantity is considered in Ref. \cite{beale}, and it is discussed in detail in the appendix \ref{app:logical}.
$\overline{\mathcal{N}}(S)_{0,0}$ is contributed only by the logical diagonal coefficients in \eqref{eq:density-matrix} and $\overline{\mathcal{N}}(S)_{0,L}$ with $[L]\neq [0]$ is contributed only by the logical off-diagonal coefficients. Both $\overline{\mathcal{N}}(S)_{0,0}$ and $\sum_{[L]}\overline{\mathcal{N}}(S)_{0,L}$ can be checked to be positive real number.
Thus we find that coherent error satisfies Eq. \ref{eq:gKL} of generalized K-L criterion,
which means toric code with single qubit coherent error is approximate quantum error correction.  
$\lambda_{S,S'}$ comes from logical diagonal terms.
$B_{S,S'}$ is proportional to the weighted summation of logical operator $X_L$, whose coefficient comes from logical off-diagonal term. We have not determined the quantity $\Delta(\Lambda+\mathcal{B},\Lambda)$ yet. But if the logical off-diagonal terms are much smaller than logical diagonal terms,
then $\Delta(\Lambda+\mathcal{B},\Lambda)$ will also be small and we can find a small number $\epsilon$ that satisfies $\Delta(\Lambda+\mathcal{B},\Lambda) \leq \epsilon$. In that case, coherent error is $\epsilon$-correctable. In fact, we can obtain a lower bound of $\Delta(\Lambda+\mathcal{B},\Lambda)$.
From \eqref{eq:lambdaB} we know that:
\begin{equation}
		\Lambda(\rho) = \sum_S \lambda_{S,S} \Tr(\rho) \ket{S}\bra{S}, \qquad
		\mathcal{B}(\rho) = \sum_S \Tr(\rho B_{S,S}) \ket{S}\bra{S}.
\end{equation}
Since $F(\Lambda+\mathcal{B},\Lambda) = \min_\rho F_\rho(\Lambda+\mathcal{B},\Lambda)$, if we set the constrain that $\rho$ only varies in the set of pure state, it will lead to an upper bound of $F$:
\begin{equation}
		\tilde{F}(\Lambda+\mathcal{B},\Lambda) = \min_{\rho,\text{ pure state}} F_\rho(\Lambda+\mathcal{B},\Lambda).
\end{equation}
and therefore, we have $F(\Lambda+\mathcal{B},\Lambda) \leq\tilde{F}(\Lambda+\mathcal{B},\Lambda)$. For pure state $\rho$, we have:
\begin{align}
		F_\rho(\Lambda+\mathcal{B},\Lambda) &= f(\frac{(\Lambda+\mathcal{B})(\rho)}{\Tr[(\Lambda+\mathcal{B})(\rho)]},\frac{\Lambda(\rho)}{\Tr[\Lambda(\rho)]})
		= \Tr \sqrt{\sqrt{\frac{(\Lambda+\mathcal{B})(\rho)}{\Tr[(\Lambda+\mathcal{B})(\rho)]}}\frac{\Lambda(\rho)}{\Tr[\Lambda(\rho)]}\sqrt{\frac{(\Lambda+\mathcal{B})(\rho)}{\Tr[(\Lambda+\mathcal{B})(\rho)]}}}\nonumber\\
		&= \frac{\sum_S \sqrt{(\lambda_{S,S}+ \Tr(\rho B_{S,S})) \lambda_{S,S}}}{\sqrt{\sum_S \lambda_{S,S}+\Tr(\rho B_{S,S})}\sqrt{\sum_{S} \lambda_{S,S}}}.
\end{align}
Note that when computing $f(\rho,\sigma)$, $\rho$ and $\sigma$ have to be normalized. 
Choose $\rho$ to be $\ket{S=0,++}$, the $+1$ eigenstate of logical Pauli $X$ operators, we have:
\begin{equation}
	\tilde{F}(\Lambda+\mathcal{B},\Lambda) \leq \frac{\sum_S \Pr(S) \sqrt{(\sum_{[L]} \overline{\mathcal{N}}(S)_{0,L}) \overline{\mathcal{N}}(S)_{0,0}}}{\sqrt{\sum_S \Pr(S) (\sum_{[L]} \overline{\mathcal{N}}(S)_{0,L})} \sqrt{\sum_S \Pr(S) \overline{\mathcal{N}}(S)_{0,0}}}
\end{equation}
So the lower bound of $\Delta$ is:
\begin{equation}
		\Delta(\Lambda+\mathcal{B},\Lambda) 
		\geq \sqrt{1-\frac{\sum_S \Pr(S) \sqrt{(\sum_{[L]} \overline{\mathcal{N}}(S)_{0,L}) \overline{\mathcal{N}}(S)_{0,0}}}{\sqrt{\sum_S \Pr(S) (\sum_{[L]} \overline{\mathcal{N}}(S)_{0,L})} \sqrt{\sum_S \Pr(S) \overline{\mathcal{N}}(S)_{0,0}}}}.
\end{equation}
We denote the lower bound as $\Delta_{lb}$. In the expression for $\Delta_{lb}$
\begin{equation}
    \begin{aligned}
        &\Delta_{lb}= \sqrt{1-\frac{\sum_S \Pr(S) \sqrt{(\overline{\mathcal{N}}(S)_{0,0} + \sum_{[L]\neq[0]} \overline{\mathcal{N}}(S)_{0,L}) \overline{\mathcal{N}}(S)_{0,0}}}{\sqrt{\sum_S \Pr(S) (\overline{\mathcal{N}}(S)_{0,0} + \sum_{[L]\neq[0]} \overline{\mathcal{N}}(S)_{0,L})} \sqrt{\sum_S \Pr(S) \overline{\mathcal{N}}(S)_{0,0}}}},
    \end{aligned}
\end{equation}
\end{widetext}
it contains both logical diagonal term $\overline{\mathcal{N}}(S)_{0,0}$ and logical off-diagonal terms $ \overline{\mathcal{N}}(S)_{0,L},[L]\neq [0]$. If the logical off-diagonal terms are small compared with the logical diagonal term, the lower bound $\Delta_{lb}$ approaches $0$. The larger the logical off-diagonal terms are, the larger $\Delta_{lb}$ becomes.  
This results indicate that if the code is $\epsilon$-correctable, the value of $\epsilon$ that describes the accuracy of the approximate or generalized QEC cannot be smaller than this lower bound. 
And due to the form of the lower bound, if the logical off-diagonal terms are large, the corresponding $\epsilon$ also has to be large.

By applying the theorem proved in Ref. \cite{beale} (see appendix \ref{app:logical}), we find that for two different error classes $[L]$, $[L']$,
\begin{equation}
	\frac{\sum_S \Pr(S) \overline{\mathcal{N}}(S)_{L,L'}}{\sum_S \Pr(S) \overline{\mathcal{N}}(S)_{0,0}} \in \mathcal{O}(\theta^{d}) \quad\text{as}\quad \theta \rightarrow 0.
\end{equation}
It means that for when the noise parameter $\theta$ is small, syndrome averaged logical off-diagonal part is smaller than syndrome averaged logical diagonal terms and scales as $\theta^d$. In other words, for fixed $\theta$, as long as the code distance $d$ approaches infinity, the logical off-diagonal terms are negligible compared with logical diagonal terms. Because from \eqref{eq:lambdaB1} we see that $\lambda_{S,S'}$ describes the logical diagonal terms and $B_{S,S'}$ describes the logical off-diagonal terms, we expect that $B_{S,S'}$ will also become ignorable compared with $\lambda_{S,S'}$ when $d\rightarrow \infty$, and the distance $\Delta(\Lambda+\mathcal{B},\Lambda)$ between the two channels $\Lambda+\mathcal{B}$ and $\Lambda$ approaches $0$ accordingly. In that case, the generalized K-L criterion \eqref{eq:gKL} approaches the normal K-L criterion \eqref{eq:KL}.

\section{Compare Coherent Error with Stochastic Error}\label{sec:compare}
The logical off-diagonal terms due to the coherent error are important quantities, which are exponentially suppressed as increasing code distance $d$~\cite{beale} and also tells us how does the generalized K-L criterion Eq.\eqref{eq:gKL} reduces to the normal K-L criterion Eq.\eqref{eq:KL} as shown in Sec.\ref{sec:KL}. However, the coherent error may change the logical diagonal term and thus the success probability of QEC. Therefore, in this section, we focus on the logical diagonal terms, and perform an analytic study to compare the effects of coherent error with the stochastic errors in the performance of the success probability. As discussed in Sec.\ref{sec:distinguish}, the error channel can be divided into two parts: the physical diagonal terms (stochastic parts after Pauli twirl) and the physical off-diagonal (coherent parts) terms as shown in the discussion after Eq.\eqref{eq:dm}. Ignoring physical off-diagonal part, which means Pauli twirl approximation, will overestimate the success probability.

We consider the probability of the error class $[E]$ under the condition of syndrome measurement outcome $S$ (defined in Sec.\ref{subsec:probability})
\begin{equation}
	\mathrm{Pr}([E]|S) = |N_S\sum_{E\in[E]} a_E|^2,
\end{equation}
which captures the information of logical diagonal part.
Set $x= \tan \theta$, we can rewrite the above expression as polynomial $x$:
\begin{equation}
	\begin{aligned}
		&\mathrm{Pr}([E]|S) = |N_S\sum_{E\in[E_c]} a_E|^2\\
		&= |N_S|^2 (\cos\theta)^{2\mathcal{E}} \sum_{c,c'\in[E]} \imi^{|c|+3|c'|} x^{|c|+|c'|}\\
		&= |N_S|^2 (\cos\theta)^{2\mathcal{E}} \sum_{b,b'\in B_1} \imi^{|E+b|+3|E+b'|} x^{|E+b|+|E+b'|},
	\end{aligned}
\end{equation}
where $\mathcal{E}$ is the number of edges, $c$ and $c'$ are the error chains belong to $[E]$, $b$ and $b'$ are boundaries ($b,b'\in [0]$), and the detail definitions for those quantities can be found in Sec.\ref{sec:procedure}. $|c|$ represents the number of edges in chain $c$. On the torus, $|E+b|$ has to be $|E|$ plus a even number (denoted as $\tilde{b}$, $|E+b|=|E|+\tilde{b}$), therefore $|E+b|+3|E+b'|$ must be even ($|E+b|+3|E+b'|= 4E + \tilde{b}+3\tilde{b'}$). Then, we obtain
\begin{align}
		&\mathrm{Pr}([E]|S)
		= |N_S|^2 (\cos\theta)^{2\mathcal{E}} \sum_{c,c'\in[E]} (-1)^{\frac{|c|+3|c'|}{2}} x^{|c|+|c'|}\nonumber\\
		&= |N_S|^2 (\cos\theta)^{2\mathcal{E}}
		\sum_{b,b'\in B_1} (-1)^{\frac{|E+b|+3|E+b'|}{2}} x^{|E+b|+|E+b'|}.
		\label{eq:chain-sum}
\end{align}
In this expression $\mathrm{Pr}([E]|S)$, the physical diagonal contribution
corresponds to $b=b'$ terms and the physical off-diagonal contribution corresponds to $b\neq b'$ terms. We set the summation part as 
\begin{equation}
 Z(x,[E]) = \sum_{c,c'\in[E]} (-1)^{\frac{|c|+3|c'|}{2}} x^{|c|+|c'|}.
\end{equation}
where $Z(x,[E])$ is proportional to $\Pr(S)\mathrm{Pr}([E]|S)$,
\begin{equation}
    \mathrm{Pr}([E]|S) = \frac{1}{\Pr(S)}(\cos\theta)^{2\mathcal{E}} Z(x,[E])
    \label{eq:unnormalized-prob}
\end{equation}
In order to perform the comparison study, we divide $Z(x,[E])$ into two parts
\begin{equation}
	\begin{aligned}
		&D(x,[E]) = \sum_{c\in[E]} x^{2|c|},\\
		&O(x,[E]) = \sum_{c\neq c'\in[E]} (-1)^{\frac{|c|+3|c'|}{2}} x^{|c|+|c'|},
	\end{aligned}
	\label{eq:zdo}
\end{equation}
where $D(x,[E])$ is the diagonal (stochastic or with Pauli twirling) part of $Z(x,[E])$ and $O(x,[E])$ is the off-diagonal (coherent) part
of $Z(x,[E])$. 
Note that $|c|+|c'| = 2|E|+\tilde{b}+\tilde{b'}$ has to be an even number, so $Z(x,[E])$ contains only even power of $x$.
We will compare $D(x,[E])$ with $O(x,[E])$.

First we focus on the stochastic part. Denote the minimum length of the elements in $[E]$ as $l_{min}$, then the length of any elements in
$[E]$ must differ from $l_{min}$ by an even number. Then $D(x,[E])$ can be reorganised by the power of $x$:
\begin{equation}
	D(x,[E]) = \sum_{n\geq 0} d_n x^{2(l_{min} + 2n)},
	\label{eq:Dpoly}
\end{equation}
where $d_n$ is the number of elements that have the length $l_{min}+2n$, we call this degeneracy. Then, we consider the calculation of $Z(x,[E])$. Setting $|c| = l_{min}+2n$, $|c'| = l_{min} + 2n'$, we get:
\begin{equation}
	\begin{aligned}
		&Z(x,[E]) = \sum_{n,n'} d_n d_{n'} (-1)^{\frac{4l_{min} + 2n + 6n'}{2}} x^{2l_{min} + 2n + 2n'} \\
		&= \sum_{m} z_m (-1)^m x^{2l_{min} + 2m},
	\end{aligned}
	\label{eq:Zpoly}
\end{equation}
where we choose the relation $m = n+n'$ in the second line above, and $z_m$ is defined as:
\begin{equation}
	z_m = \sum_{n,n'|n+n'=m} d_n d_{n'}.
	\label{eq:zm}
\end{equation}
In this case, the coherent part $O(x,[E])$ can be obtain from $Z(x,[E])-D(x,[E])$, so we have:
\begin{equation}
	O(x,[E]) = \sum_{m} o_m (-1)^m x^{2l_{min} + 2m},
\end{equation}
where
\begin{equation}
	o_m =  \left\{
	\begin{aligned}
	z_m-d_{m/2}, \quad m\mod{2} = 0\\
	z_m, \quad m\mod{2} = 1,\\
	\end{aligned}
	\right.
	\label{eq:om}
\end{equation}
where $z_m$ is defined in Eq.\eqref{eq:zm} and $d_m$ is defined in Eq.\eqref{eq:Dpoly} for the reminder.

With the expression of $O(x,[E])$ and $D(x,[E])$, we can compare them under small noise limit ($\theta$ and thus $x$ is small) by a Taylor expansion at $x\rightarrow 0$:
\begin{equation}
		\frac{O(x,[E])}{D(x,[E])} = \frac{\sum_{m} o_m (-1)^m x^{2m}}{\sum_{n} d_n x^{4n}}
		= \sum_{m} r_m x^{2m}
\end{equation}
Using Eq.\eqref{eq:zm} and \eqref{eq:om}, the lowest order terms can be computed explicitly:
\begin{equation}
	\begin{aligned}
		&r_0 = \frac{z_0-d_0}{d_0} = d_0 - 1,\\
		&r_1 = -\frac{z_1}{d_1} = -2d_1.
	\end{aligned}
\end{equation}
So we get the ratio of coherent part and stochastic part under small noise limit:
\begin{equation}
	\begin{aligned}
		&\frac{O(x,[E])}{D(x,[E])} \rightarrow (d_0-1) - 2d_1 x^2 + r_2 x^4 + \cdots\\
		&= (d_0-1) - 2d_1 (\tan\theta)^2 + r_2 (\tan\theta)^4 + \cdots.
	\end{aligned}
	\label{eq:ratio}
\end{equation}
For examples, we may consider the following three kinds of situations.
In these examples, we assume the number of edges $\mathcal{E}$, which is also the number of qubit since each edge is assigned with a qubit, is large enough, and the error parameter $x=\tan\theta$ is small, $x \rightarrow 0$. 

\begin{figure}[t]
	\centering
	\includegraphics[scale=0.4]{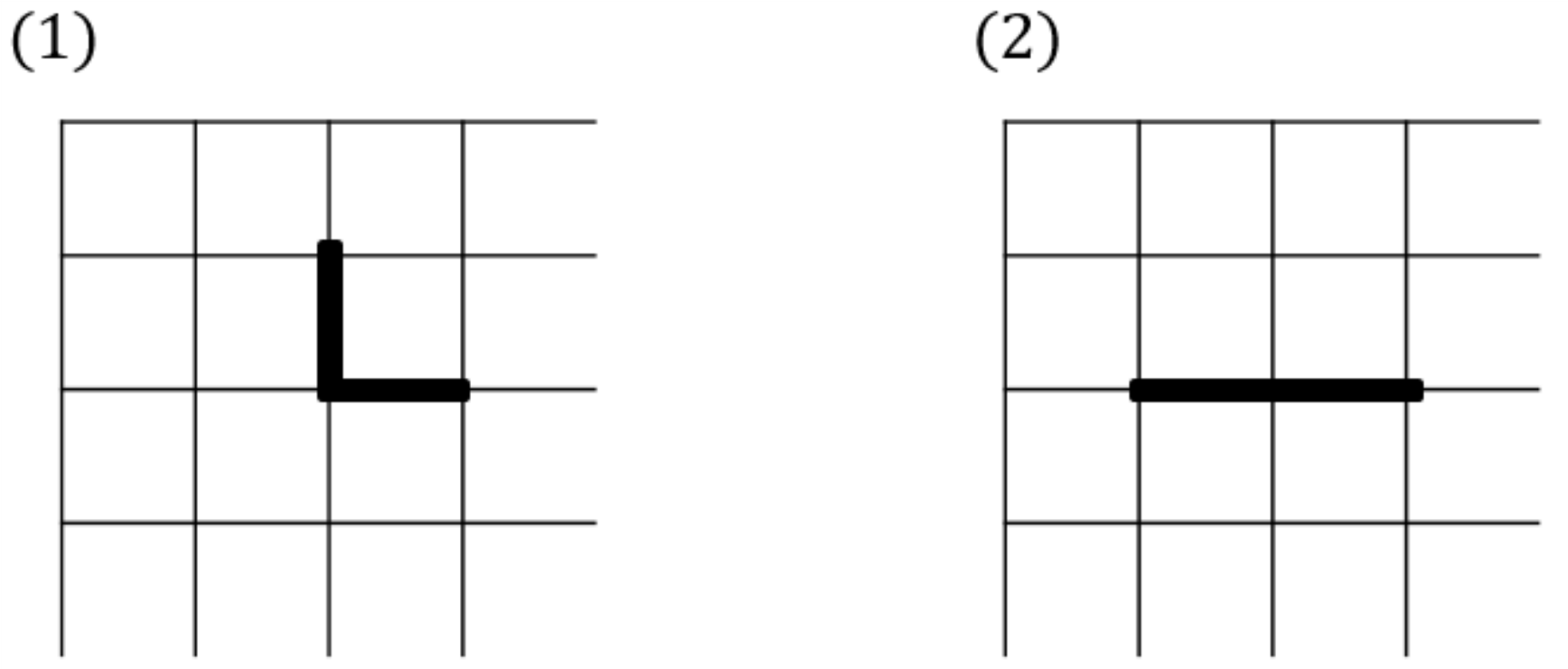}
	\caption{Two error classes with different degeneracy.}
	\label{fig:ratio-example}
\end{figure}

First of all, we consider the case $[E]=[0]$, which means $[E]$ is the class of error chains that form boundaries of regions and can be viewed as $X$ stabilizers. In the error chain class $[0]$, the shortest error chain is just empty error chain $E=0$, so $l_{min}=0$, and there is only one element with length $0$, so $d_0 = 1$. There are no elements with length $2$ in class $[0]$, since length $2$ error chains cannot enclose a region, so $d_1 = 0$. As for the error chains with length $4$, each of these error chains just enclose a single face. Since the number of faces is half of the number of edges, we have $d_2 = \mathcal{E}/2$. 
Therefore, the ratio of coherent part and stochastic part becomes
\begin{equation}
	\frac{O(x,[E])}{D(x,[E])} \rightarrow \mathcal{E} x^4 + \cdots.
	\label{eq:example1}
\end{equation} 
If the noise strength is small enough, the coherent part can be ignored compared with stochastic part.

For the other two examples, we consider the two error classes in FIG. \ref{fig:ratio-example}. For error class (1), the error chains in this class can be any 'continuous' deformation of the error chain in (1) of FIG.  \ref{fig:ratio-example}, which means the difference between them is a boundary of regions. The shortest length in this class is obviously $l_{min}=2$. By counting the number of error chains that have the length $2$, $4$ and $6$ similarly as before, we obtain that the degeneracy $d_n$ for length $l_{min}+2n$ error chains when $n=0,1,2$ are $d_0 = 2$, $d_1 = 4$ and $d_2 = \mathcal{E} - 4$. Substitute these results into Eq. \eqref{eq:zm} and \eqref{eq:om}, we obtain $z_0 = 4$, $z_1 = 16$, $z_2 = 4\mathcal{E}$, $o_0 = 2$, $o_1 = 16$, $o_2 = 4\mathcal{E} - 4$, and the ratio of coherent part and stochastic part is:
\begin{equation}
	\begin{aligned}
		&\frac{O(x,[E])}{D(x,[E])} = \frac{2-16x^2 + (4\mathcal{E}-4)x^4 + \cdots}{2+ 4 x^4 + \cdots}\\
		&= 1-8x^2 + (2\mathcal{E}-4)x^4+\cdots.
	\end{aligned}
	\label{eq:example2}
\end{equation}
Note that in this case, even for very small but finite $x$, the ratio ${O(x,[E])}/{D(x,[E])}$ does not vanish, but acquire a constant value. This result is not inconsistent with the situation at $x=0$ limit, because this error class will first vanish in the $x=0$ limit.
This means that even when the noise is small but finite, the coherent part could be comparable to the stochastic part for certain degenerate error chains.
For error class (2), we do the counting similarly and find that $d_0 = 1$, $d_1 = 6$, $d_2 = \mathcal{E}/2 - 2$. Substitute these results into Eq. \eqref{eq:zm} and \eqref{eq:om}, we obtain $z_0 = 1$, $z_1 = 12$, $z_2 = \mathcal{E} + 32$, $o_0 = 0$, $o_1 = 12$, $o_2 = \mathcal{E} + 26$, then the ratio of coherent part and stochastic part is:
\begin{equation}
	\begin{aligned}
		&\frac{O(x,[E])}{D(x,[E])} = \frac{-12x^2 + (\mathcal{E}+26)x^4 + \cdots}{1+ 6 x^4 + \cdots}\\
		&= -12x^2 + (\mathcal{E}+26)x^4+\cdots
	\end{aligned}
	\label{eq:example3}
\end{equation}
In this case the leading term scales as $x^2$. The above three examples tells us that while the noise is small, the coherent part could be comparable to stochastic part for some error classes, and is negligible in other cases.

\begin{figure}[t]
	\centering
	\includegraphics[scale=0.25]{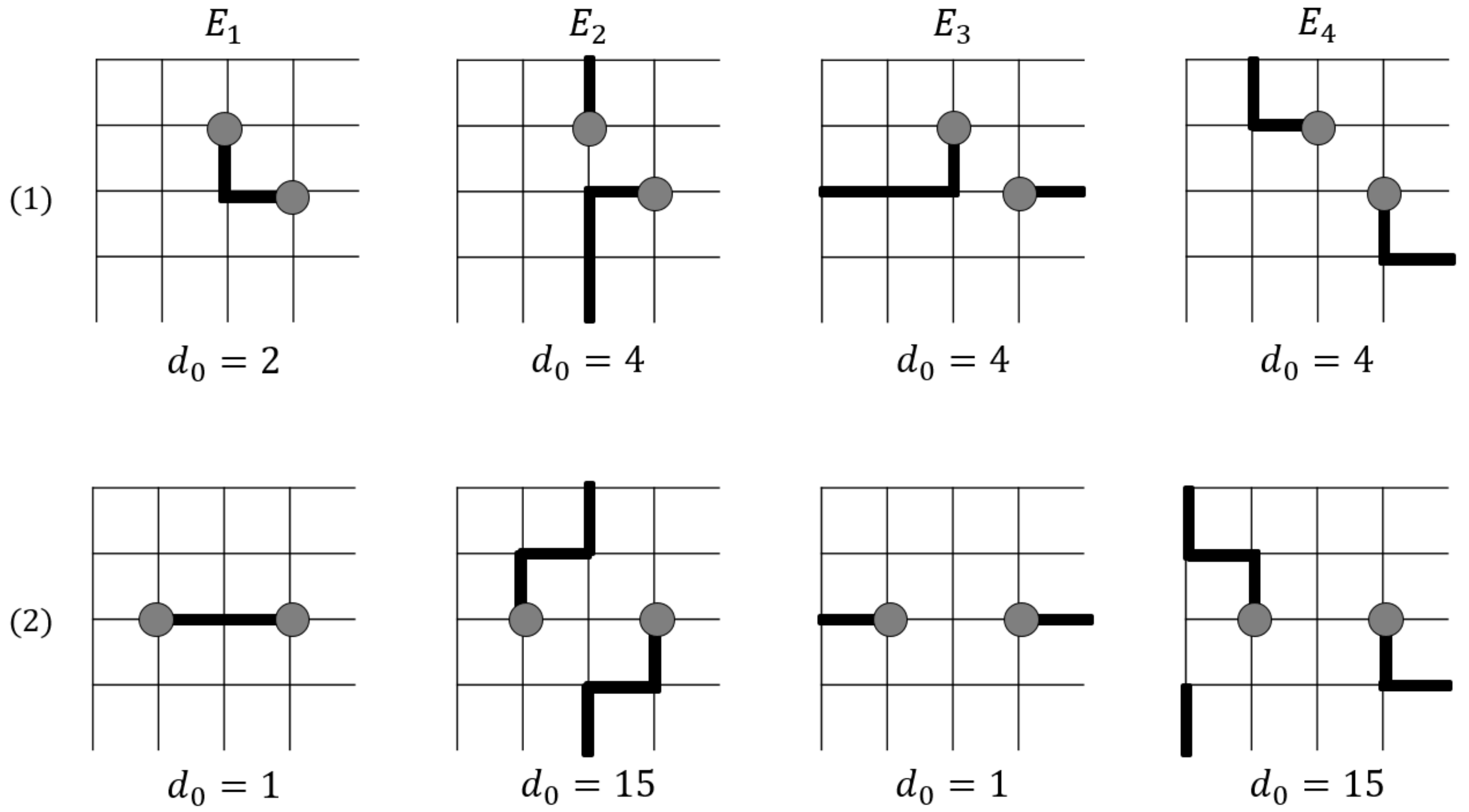}
	\caption{Two syndromes and the shortest error chain in four error classes correspond to each syndrome on $4\times 4$ lattice}
	\label{fig:suc-rate}
\end{figure}

So far from Eq. \eqref{eq:ratio} we see that when degeneracy $d_0$ is larger than $1$, the coherent part can reach a constant multiple of the stochastic part while noise is small.
Now the question is how does those degenerate error chains affect the error correction procedure.
When the noise $\theta$ is small, only the shortest error chain has the largest contribution in the error chain summation of $Z(x,[E])$. So in that case, the maximum likelihood decoder is equivalent to minimum weight perfect matching decoder, which choose out the error class that contains the shortest error chain (this error class has the largest probability). For a given syndrome $S$, denote the four different error classes of this syndrome as $[E_{S,i}]$, $i=1,2,3,4$.
Set $[E_{S,1}]$ as the error class with the largest probability, and it is chosen by decoder as the error correction that we should apply, then the success probability of this error correction is $\mathrm{Pr}([E_{S,1}]|S)$ as discussed in subsection \ref{subsec:probability}, and it is expressed in $Z(x,[E_{S,i}])$ as:
\begin{equation}
	\mathrm{Pr}([E_{S,1}]|S)  = \frac{Z(x,[E_{S,1}])}{\sum_i Z(x,[E_{S,i}])}.
\end{equation}
Here, $Z(x,[E_{S,i}])$ is unnormalized probability of error class $[E_{S,i}]$, and the success probability is obtained by dividing a normalization factor $\sum_i Z(x,[E_{S,i}])$ under the particular syndrome $S$.
We denote:
\begin{equation}
	R_i = \frac{Z(x,[E_{S,i}])}{D(x,[E_{S,i}])} = \frac{O(x,[E_{S,i}])}{D(x,[E_{S,i}])}+1,
\end{equation}
then the success probability can be written as
\begin{equation}
	\begin{aligned}
		&\mathrm{Pr}([E_{S,1}]|S) = \frac{R_1 D(x,[E_{S,1}])}{\sum_i R_i D(x,[E_{S,i}])}\\
		&= \frac{D(x,[E_{S,1}])}{D(x,[E_{S,1}]) +\sum_{i=2}^{4} (R_i/R_1) D(x,[E_{S,i}])}.
	\end{aligned}
\label{eq:suc-rate}
\end{equation} 
Similarly, if we consider stochastic error under Pauli twirl approximation instead of coherent error, the success probability becomes
\begin{equation}
	\mathrm{Pr}_{twirl}([E_{S,1}]|S) = \frac{D(x,[E_{S,1}])}{\sum_i D(x,[E_{S,i}])}
\label{eq:sto-suc-rate}
\end{equation}
Note that under small noise, the maximum likelihood decoder that finds for the most possible error class for coherent error and for stochastic error will both choose out the same error class of a syndrome with the shortest error chain in it, Here $[E_{S,1}]$, since this error class will have the largest probability in both cases, that is why $\mathrm{Pr}_{twirl}([E_{S,1}]|S)$ will be viewed as success probability for stochastic error.
When the noise is small, $x \rightarrow 0$, we know that $R_i/R_1 \simeq d_{0,[E_{S,i}]}/d_{0,[E_{S,1}]}$, where $d_{0,[E_{S,i}]}$ is the degeneracy of shortest error chain $d_0$ in the error class $[E_{S,i}]$.
Because error class $[E_{S,1}]$ has the shortest error chain length, and smaller error chain is more likely to have smaller degeneracy $d_0$ (for example in FIG. \ref{fig:suc-rate}),
we have $d_{0,[E_{S,i}]}/d_{0,[E_{S,1}]} \geq 1$ for $i=2,3,4$.
So the denominator of Eq. \eqref{eq:suc-rate} is larger than that in Eq. \eqref{eq:sto-suc-rate},
and we have the conclusion that the coherency of error reduces the success probability of error correction.

We now compare the success probability $\mathrm{Pr}([E_{S,1}]|S)$ for coherent error and $\mathrm{Pr}_{twirl}([E_{S,1}]|S)$ for stochastic error quantitatively.
From Eq. \eqref{eq:suc-rate} and \eqref{eq:sto-suc-rate} we obtain:
\begin{align}
&\frac{1}{\mathrm{Pr}([E_{S,1}]|S)} - \frac{1}{\mathrm{Pr}_{twirl}([E_{S,1}]|S)} \nonumber\\
    &\qquad\qquad\qquad= \sum_{i=2}^{4} (\frac{R_i}{R_1} - 1) \frac{D(x,[E_{S,i}])}{D(x,[E_{S,1}])},
\end{align}
and consequently we obtain the ratio of the success probability for stochastic parts and coherent parts
\begin{equation}
	\begin{aligned}
		&\frac{\mathrm{Pr}_{twirl}([E_{S,1}]|S)}{\mathrm{Pr}([E_{S,1}]|S)}\\
		& = 1+ \mathrm{Pr}_{twirl}([E_{S,1}]|S)\sum_{i=2}^{4} (\frac{R_i}{R_1} - 1)\frac{D(x,[E_{S,i}])}{D(x,[E_{S,1}])}\\
		&= 1+ \sum_{i=2}^{4} (\frac{R_i}{R_1} - 1)\frac{D(x,[E_{S,i}])}{\sum_{j=1}^{4} D(x,[E_j])}.
	\end{aligned}        
	\label{eq:success-probability1}
\end{equation}
For small noise $x \rightarrow 0$, we want to keep the lowest order of $x$ in the above expression. Note that each term in the $i$ summation might have different lowest order because
of the difference of $l_{min,[E_{S,i}]}$.
Because error class $[E_{S,1}]$ has the largest probability, we suppose that $l_{min,[E_{S,1}]} \leq l_{min,[E_{S,2}]}$, $l_{min,[E_{S,3}]}$, $l_{min,[E_{S,4}]}$.
Up to the lowest order, $\sum_{j} D(x,[E_j]) \simeq D(x,[E_{S,1}]) \simeq d_{0,[E_{S,1}]} x^{2l_{min,[E_{S,1}]}}$, $D(x,[E_{S,i}]) \simeq d_{0,[E_{S,i}]} x^{2l_{min,[E_{S,i}]}}$.
Then we retain the lowest order in the $i$ summation, and get:
\begin{equation}
    \begin{aligned}
    	&\frac{\mathrm{Pr}_{twirl}([E_{S,1}]|S)}{\mathrm{Pr}([E_{S,1}]|S)}=\\
    	&1+ \sum_{\substack{i=2 \\ \text{lowest}\\ \text{order}}}^{4} \frac{d_{0,[E_{S,i}]}}{d_{0,[E_{S,1}]}}(\frac{d_{0,[E_{S,i}]}}{d_{0,[E_{S,1}]}} - 1) x^{2(l_{min,[E_{S,i}]} - l_{min,[E_{S,1}]})}.
    \end{aligned}
	\label{eq:success-probability2}
\end{equation}
Here the lowest order of $x$ is retained in the summation. 
Clearly the above expression is greater than or equal to $1$ because ${d_{0,[E_{S,i}]}}/{d_{0,[E_{S,1}]}} - 1\geq 0$ and for most cases it is strictly greater than $1$ since the degeneracy of shortest error chain is generally than other error chains. So stochastic error can be corrected with a higher success rate than coherent error.
We see that as $x$ grows, ${\mathrm{Pr}_{twirl}([E_{S,1}]|S)}/{\mathrm{Pr}([E_{S,1}]|S)}$ becomes larger and larger than $1$, so the success probability under coherent error becomes smaller and smaller compared with stochastic error. The ratio ${\mathrm{Pr}_{twirl}([E_{S,1}]|S)}/{\mathrm{Pr}([E_{S,1}]|S)}$ equals to $1$ when $x=0$, and this is trivially obvious when no error occurs.
However, it seems to be contradictory to the fact that Eq. \eqref{eq:ratio} have nonzero constant term. However, that is not the case.
The nonzero constant term in Eq. \eqref{eq:ratio} contributes to $R_i/R_1-1$ in Eq. \eqref{eq:success-probability1}, which leads to
the part $d_{0,[E_{S,i}]}/d_{0,[E_{S,1}]} - 1$ in Eq. \eqref{eq:success-probability2}. If we assume that the constant term of $O(x,[E])/D(x,[E])$ is $0$ for all $[E]$ (though this assumption is not realistic),
then $d_{0,[E_{S,i}]}/d_{0,[E_{S,1}]} - 1$ must be zero and the second term in Eq. \eqref{eq:success-probability2} vanishes.
Off that situation, only higher order terms of $x$ are left. So the nonzero constant term in Eq. \eqref{eq:ratio} is the reason why the success probability of coherent error is smaller than
Pauli twirled stochastic error when noise is small.

\section{Summary and Discussion}\label{sec:con}

 In this paper, we study focus on the QEC procedure of the toric code, and consider the error correction performance the independent coherent error. We find that the toric code under this coherent error model can be captured by the generalized Knill-Laflamme (K-L) criterion and falls into the category of approximate QEC. The original term that that exists in the exact K-L criterion corresponds to the logical diagonal parts of the error channel, while the extra term in the generalized K-L criterion corresponds to the coherent part of the error channel at logical level. We then show that the generalized K-L criterion approaches the normal K-L criterion when the code distance becomes large. In addition, we also find that if the code with a fixed distance $d$ is $\epsilon$- correctable, the value of $\epsilon$ describing the accuracy of the approximate QEC cannot be smaller than a lower bound. We then study the success probability of QEC under such coherent errors, and confirm that the exact success probability under coherent error is smaller than the results using Pauli twirling approximation at physical level. 

In reality, other types of coherent errors beyond the independent coherent rotation errors, which assume perfect syndrome measurements, could be important in quantum computer. For example, detection induced coherent errors with imperfect syndrome measurements and the effect of constant qubit-qubit interactions which cannot be fully turned off, e.g. ZZ crosstalk in CZ gate of SC qubits. One of the authors analyzed the effect of detection-induced coherent errors introduced by controlled-Z gates for the surface code~\cite{yang2021quantum}. The results show that the detection-induced coherent error will result in undetected error terms, and they will evolve into logical errors after several QEC cycles, and the observation indicates that although those undetected error terms may be less harmful for one QEC cycle, after several QEC cycles they will evolve into logical errors and become more harmful~\cite{yang2021quantum}. Only if we fixed the number of QEC cycles and the small coherent rotation angle, one can show that the surface code can alleviate this effect by increasing the code size, and thus demonstrate the effectiveness of QEC again similar to the results for the independent coherent rotation errors~\cite{Flammia-PRA,beale,Iverson_2020}. Finally, we want to emphasize that the general coherent noise problem for QEC and fault-tolerant quantum computation is still unsolved and in its early stage of research.



\begin{acknowledgments} 
We thank Qinghong Yang, Li Rao and Guangming Zhang for helpful discussions. The work is supported by Natural Science Foundation of China (Grants No.~11974198) and Tsinghua University Initiative Scientific Research Program.
\end{acknowledgments}

\begin{appendix}
\begin{widetext}

\section{Logical Diagonal and Off-diagonal Part}\label{app:logical}
In this part, we apply the theorem proved in Ref.\cite{beale} to our model.
In Ref.\cite{beale}, they considered the quantity 
\begin{equation}
	\begin{aligned}
		&\overline{\mathcal{N}}(S)_{L,L'} = \frac{\langle\bra{P_L\Pi_S} \mathcal{N} \ket{P_{L'}\Pi_0}\rangle}{\Pr(S)2^k}\\
		&\mathcal{N} = \bigotimes_{i=1}^n \mathcal{N}^{(i)}
	\end{aligned}
\end{equation}
for a general $[[n,k,d]]$ stabilizer code and proved that
\begin{equation}
	\begin{aligned}
		&\sum_s \Pr(S) \overline{\mathcal{N}}(S)_{L,L'} \in \mathcal{O}(r^{d/2}) \quad\text{as}\quad r \rightarrow 0\\
		&r = \max_i \Tr(I-\mathcal{N}^{(i)})/6.
	\end{aligned}        
\end{equation}
Here $r$ is the average infidelity of a single-qubit noise process, $\Pi_S$ is the projector onto syndrome $S$, $P_L,P_{L'}$ are any two logical Pauli operators of the stabilizer code labeled by $L$ and $L'$, $\mathcal{N}^{(i)}$ is the error channel on a single qubit $i$, $\mathcal{N}$ is the overall error channel.
The double vector $\ket{\cdot}\rangle$ is defined as follows.
For a $m$-dimensional system, $\mathbb{C}^{m\times m}$ is the space of operators on this system. Let $\mathbb{B}$ be an arbitrary trace-orthonormal basis of $\mathbb{C}^{m\times m}$, $\Tr(B_j^\dagger B_k)=\delta_{j,k}$ for all $B_j, B_k\in \mathbb{B}$. For example $\mathbb{B}$ can be chosen as Pauli basis. We use double vector to represent operators in $\mathbb{C}^{m\times m}$ as vectors. The base vector $B_j\in \mathbb{B}$ is expressed as $\ket{B_j}\rangle$. For any $M\in \mathbb{C}^{m\times m}$, $\ket{M}\rangle=\sum_j \Tr(B_j^\dagger M)\ket{B_j}\rangle$. The inner product is $\langle \braket{M|N} \rangle=\Tr(M^\dagger N)$. For a quantum channel $\mathbb{N}$ that maps a system to another, its action on any vector $\ket{M}\rangle$ is defined as $\mathcal{N}\ket{M}\rangle=\ket{\mathcal{N}(M)}\rangle=\sum_{B\in \mathbb{B}_{in}}\ket{\mathcal{N}(B)}\rangle\langle \braket{B|M} \rangle$, where $\mathbb{B}_{in}$ is the trace-orthonormal basis for the input system.

In our error correction procedure, the error channel is the single qubit unitary $X$ rotation $\mathcal{N}(\rho) = U\rho U^\dagger$, and logical qubit number is $k=2$. We choose $P_L,P_{L'}$ to be logical Pauli $X$ operators $X_L$ and $X_{L'}$, and $L,L'$ are the logical chains.
We denote the states of our system as $\ket{S,x}$, where $S$ is syndrome and $x$ is a two bit binary number that stands for logical degree of freedom ($\ket{S,x}=X_{E_{S,max}} \ket{0,x}$, $X_{E_{S,max}}$ has maximum probability under syndrome $S$).
Then the projector $\Pi_S$ is:
\begin{equation}
	\Pi_S = \sum_{x} \ket{S,x}\bra{S,x}.
\end{equation}
Then we obtain:
\begin{equation}
	\begin{aligned}
		&\overline{\mathcal{N}}(S)_{L,L'} = \frac{1}{4\Pr(S)}\Tr(\Pi_S X_L U X_{L'} \Pi_0 U^\dagger)\\
		&= \frac{1}{4\Pr(S)}\Tr((\sum_{x} \ket{S,x}\bra{S,x}) X_L  X_{L'}\sum_{E,E'} a_E a_{E'}^* X_E (\sum_y \ket{0,y}\bra{0,y}) X_{E'})\\
		&= \frac{1}{4\Pr(S)}\sum_{x,y}\sum_{E,E'} \bra{S,x} a_E a_{E'}^* X_{E+L+L'} \ket{0,y}\bra{0,y} X_{E'} \ket{S,x}\\
		&= \frac{1}{4\Pr(S)}\sum_{x,y}\sum_{E|\partial E =S,E'|\partial E' =S}  a_E a_{E'}^*\bra{S,x} X_{E+L+L'} \ket{0,y}\bra{0,y} X_{E'} \ket{S,x}.
	\end{aligned}
\end{equation}
As before, we split the summation of error chains under a particular syndrome $\sum_{E|\partial E =S}$ to summation of logical classes $\sum_{[J]}$ and 
summation of error chains in a logical class $\sum_{E_J \in [E_{S,max} + J]}$:
\begin{equation}
	\begin{aligned}
		&\overline{\mathcal{N}}(S)_{L,L'} = \frac{1}{4\Pr(S)}\sum_{x,y}\sum_{[J],[J']}\sum_{E_J \in [E_{S,max} + J]}\sum_{E_{J'} \in [E_{S,max} + J']}  a_{E_J} a_{E_{J'}}^* \bra{S,x} X_{E_J+L+L'} \ket{0,y}\bra{0,y} X_{E_{J'}} \ket{S,x}\\
		&= \frac{1}{4\Pr(S)}\sum_{x,y}\sum_{[J],[J']}\sum_{E_J \in [E_{S,max} + J]}\sum_{E_{J'} \in [E_{S,max} + J']} a_{E_J} a_{E_{J'}}^* \bra{S,x} X_{J+L+L'} \ket{S,y}\bra{S,y} X_{J'} \ket{S,x}.
	\end{aligned}
\end{equation}
Let $L_x$ be the logical chain that corresponds to $x$. In the above expression, non-vanishing terms have the constrain $X_{L_x} X_{J} X_L X_{L'} X_{L_y} = I $, $X_{L_y} X_{J'} X_{L_x} = I$,
which means $[L_x] + [J] + [L] + [L'] + [L_y]= [0]$, $[L_y] + [J'] + [L_x]= [0]$.
So we obtain:
\begin{equation}
	\begin{aligned}
		&\overline{\mathcal{N}}(S)_{L,L'} = \frac{1}{\Pr(S)}\sum_{[J]}\sum_{E \in [E_{S,max} + J + L  + L']}  \sum_{E' \in [E_{S,max} + J]}a_{E} a_{E'}^*\\
		&=\frac{1}{\Pr(S)}\sum_{[J]}\sum_{E \in [E_{S,max} + J + L]}  \sum_{E' \in [E_{S,max} + J + L']}a_{E} a_{E'}^*\\
		&\sum_S \Pr(S) \overline{\mathcal{N}}(S)_{L,L'} = \sum_S \sum_{[J]}\sum_{E \in [E_{S,max} + J + L]}  \sum_{E' \in [E_{S,max} + J + L']}a_{E} a_{E'}^*.
	\end{aligned}
	\label{eq:logical-coefficient}
\end{equation}
This is the quantity considered in Ref. \cite{beale}. The $L=L'$ logical diagonal terms of $\overline{\mathcal{N}}(S)_{L,L'}$ is just the summation of the coefficients of
coherent error logical diagonal term in equation \eqref{eq:density-matrix}, same as the off-diagonal terms.
The quantity $\lambda_{S,S'}$ and $B_{S,S'}$ in section \ref{sec:KL} can be expressed by $\overline{\mathcal{N}}(S)_{L,L'}$:
\begin{equation}
	\begin{aligned}
		&\lambda_{S,S'} = \delta_{S,S'} \sum_{[L]} \sum_{E,E' \in [E_{S,max} + L]} a^*_{E'} a_{E} = \delta_{S,S'} \mathrm{Pr}(S) \overline{\mathcal{N}}(S)_{0,0}\\
		&B_{S,S'} = \delta_{S,S'} \sum_{[L] \neq [0]} X_{L} \sum_{[J]} \sum_{E_J \in [E_{S,max} + J]} \sum_{E_{J+L} \in [E_{S,max} + J+L]} a_{E_J} a^*_{E_{J+L}}\\
		&=\delta_{S,S'} \mathrm{Pr}(S) \sum_{[L] \neq [0]} X_{L} \overline{\mathcal{N}}(S)_{0,L}.
	\end{aligned}
	\label{eq:lambdaB}
\end{equation}

For our single qubit coherent error $\mathcal{N}^{(i)}(\rho) = e^{i\theta X} \rho e^{-i\theta X}$, the average infidelity $r$ can be expressed in $\theta$.
To compute $\Tr(I-\mathcal{N}^{(i)})/6$, we choose the Pauli operators $\{I,X,Y,Z\}/\sqrt{2}$ as the basis of $\rho$.
An arbitrary density matrix $\rho = a (I/\sqrt{2}) + b (X/\sqrt{2}) + c (Y/\sqrt{2}) + d(Z/\sqrt{2})$ can be written in 
vector form as $\ket{\rho}\rangle = a \ket{I/\sqrt{2}}\rangle + b \ket{X/\sqrt{2}}\rangle + c \ket{Y/\sqrt{2}}\rangle + d\ket{Z/\sqrt{2}}\rangle$.
We can compute the matrix elements of $\mathcal{N}^{(i)}$ in Pauli basis.
Since $\mathcal{N}^{(i)}(I/\sqrt{2}) = e^{i\theta X} (I/\sqrt{2}) e^{-i\theta X} = I/\sqrt{2}$, we have:
\begin{equation}
	\ket{\mathcal{N}^{(i)}(I/\sqrt{2})}\rangle = \ket{I/\sqrt{2}}\rangle.
\end{equation}
Similarly:
\begin{equation}
	\begin{aligned}
		&\ket{\mathcal{N}^{(i)}(X/\sqrt{2})}\rangle = \ket{X/\sqrt{2}}\rangle\\
		&\ket{\mathcal{N}^{(i)}(Y/\sqrt{2})}\rangle = \cos 2\theta \ket{Y/\sqrt{2}}\rangle - \sin 2\theta \ket{Z/\sqrt{2}}\rangle\\
		&\ket{\mathcal{N}^{(i)}(Z/\sqrt{2})}\rangle = \cos 2\theta \ket{Z/\sqrt{2}}\rangle + \sin 2\theta \ket{Y/\sqrt{2}}\rangle.
	\end{aligned}
\end{equation}
Thus the average infidelity is:
\begin{equation}
	r=\Tr(I-\mathcal{N}^{(i)})/6= \frac{1-\cos 2\theta}{3}.
\end{equation} 
Obviously $0\leq r \leq 2/3$. From the result of reference \cite{beale} we know that for the logical off-diagonal $L\neq L'$ term:
\begin{equation}
	\sum_S \Pr(S) \overline{\mathcal{N}}(S)_{L,L'} \in \mathcal{O}(r^{d/2}) \quad\text{as}\quad r \rightarrow 0.
\end{equation}
Expressed in $\theta$:
\begin{equation}
	\sum_S \Pr(S) \overline{\mathcal{N}}(S)_{L,L'} \in \mathcal{O}(\theta^{d}) \quad\text{as}\quad \theta \rightarrow 0.
\end{equation}

For the logical diagonal term, we have:
\begin{equation}
	\begin{aligned}
		&\sum_S \Pr(S) \overline{\mathcal{N}}(S)_{L,L} = \sum_S \Pr(S) \overline{\mathcal{N}}(S)_{0,0}\\
		&= \sum_S \sum_{[J]}\sum_{E \in [E_{S,max} + J]}  \sum_{E' \in [E_{S,max} + J]}a_{E} a_{E'}^*\\
		&= \sum_S \sum_{[J]}\Pr([J],S)
		=1
	\end{aligned}
\end{equation}
In our system, the averaged logical diagonal elements $\sum_S \Pr(S) \overline{\mathcal{N}}(S)_{L,L}$ will always be equal to $1$.
So we have the conclusion that when $\theta$ is small, the syndrome averaged logical off-diagonal elements $\sum_S \Pr(S) \overline{\mathcal{N}}(S)_{L,L'}$ decays exponentially faster that logical diagonal elements $\sum_S \Pr(S) \overline{\mathcal{N}}(S)_{L,L}$ 
with respect to code distance $d$, the ratio of them scales as $\mathcal{O}(\theta^d)$. This result implies that for small noise and large error distance $\Delta(\mathcal{\Lambda+B},\Lambda)$ approaches $0$ and the generalized K-L criterion \eqref{eq:gKL} approaches the normal K-L criterion \eqref{eq:KL}.

In addition, we note that this result justifies our choice to measure the logical error with fidelity. Generally, the diamond norm distance is needed to measure the influence of coherent error, because it intakes the logical off-diagonal contribution \cite{Flammia-PRA}, as fidelity only contains logical diagonal terms. But with the fact that logical off-diagonal terms are negligible to logical diagonal terms, we expect that the difference between diamond norm and logical infidelity is also negligible, and the effect of coherent error is well characterised by logical infidelity \cite{beale}.

\end{widetext} 
\end{appendix}

\bibliography{ref,DICE_Ref}
\bibliographystyle{apsrev4-1}

\end{document}